\documentclass[preprint,superscriptaddress,floatfix,sort&compress,twocolumn,10pt,tightenlines]{revtex4}

\usepackage[dvips]{graphicx}
\usepackage{amssymb,amsfonts,amsmath}
\usepackage{color}
\usepackage[font=scriptsize]{caption}

\usepackage[position=top]{subfig}

\usepackage{ulem}

\newcommand{\1}{\begin{equation}}
\newcommand{\2}{\end{equation}}
\newcommand{\ea}{\begin{eqnarray}} 
\newcommand{\ee}{\end{eqnarray}}
\newcommand{\4}[2]{{\frac{#1}{#2}}} 
 
\newcommand{\Sum}[2]{{\sum\limits_{#1}^{#2}}}

\newcommand{\I}{{ {\rm i}  }}

\newcommand{\SI}{\textit{Supporting Information}}

\begin{document}

\title{Ephemeral protein binding to DNA shapes stable nuclear bodies and chromatin domains}

\author{C. A. Brackley}
\affiliation{SUPA, School of Physics and Astronomy, University of Edinburgh, Peter Guthrie Road, Edinburgh, EH9 3FD, UK}
\author{B. Liebchen}
\affiliation{SUPA, School of Physics and Astronomy, University of Edinburgh, Peter Guthrie Road, Edinburgh, EH9 3FD, UK}
\author{D. Michieletto}
\affiliation{SUPA, School of Physics and Astronomy, University of Edinburgh, Peter Guthrie Road, Edinburgh, EH9 3FD, UK}
\author{F. Mouvet}
\affiliation{SUPA, School of Physics and Astronomy, University of Edinburgh, Peter Guthrie Road, Edinburgh, EH9 3FD, UK}
\author{P. R. Cook}
\affiliation{Sir William Dunn School of Pathology, University of Oxford, South Parks Road, Oxford, OX1 3RE, UK}
\author{D. Marenduzzo}
\affiliation{SUPA, School of Physics and Astronomy, University of Edinburgh, Peter Guthrie Road, Edinburgh, EH9 3FD, UK}

\begin{abstract}
\bf{Fluorescence microscopy reveals that the contents of many (membrane-free) nuclear ``bodies'' exchange rapidly with the soluble pool whilst the underlying structure persists; such observations await a satisfactory biophysical explanation. To shed light on this, we perform large-scale Brownian dynamics simulations of a chromatin fiber interacting with an ensemble of (multivalent) DNA-binding proteins; these proteins switch between two states -- active (binding) and inactive (non-binding). This system provides a model for any DNA-binding protein that can be modified post-translationally to change its affinity for DNA (e.g., like the phosphorylation of a transcription factor). Due to this out-of-equilibrium process, proteins spontaneously assemble into clusters of self-limiting size, as individual proteins in a cluster exchange with the soluble pool with kinetics like those seen in photo-bleaching experiments. This behavior contrasts sharply with that exhibited by ``equilibrium'', or non-switching, proteins that exist only in the binding state; when these bind to DNA non-specifically, they form clusters that grow indefinitely in size. Our results point to post-translational modification of chromatin-bridging proteins as a generic mechanism driving the self-assembly of highly dynamic, non-equilibrium, protein clusters with the properties of nuclear bodies. Such active modification also reshapes intra-chromatin contacts to give networks resembling those seen in topologically-associating domains, as switching markedly favors local (short-range) contacts over distant ones. }
\end{abstract}

\maketitle

\newpage

In all living organisms, from bacteria to man, DNA and chromatin are invariably associated with binding proteins, which organize their structure~\cite{understandingDNA,alberts,peter}. Many of these architectural proteins are molecular bridges that can bind at two or more distinct DNA sites to form loops. For example, bacterial DNA is looped and compacted by the histone-like protein H-NS which has two distinct DNA-binding domains~\cite{HNS}. In eukaryotes, complexes of transcription factors and RNA polymerases stabilize enhancer-promoter loops~\cite{Simonis2006,Barbieri2012,GenomeBiology,NAR}, while HP1~\cite{HP1}, histone H1~\cite{H1}, and the polycomb-repressor complex PRC1/2~\cite{Ajaz,Davide} organize inactive chromatin. Proteins also bind to specific DNA sequences to form larger structures, like nucleoli and the histone-locus, Cajal, and promyeloleukemia bodies~\cite{nuclearbodies,Sleeman2014,Mao2011,Pirrotta2012,Berry2015,Brangwynne2015}. The selective binding of molecular bridges to active and inactive regions of chromatin has also been highlighted as one possible mechanism underlying the formation of topologically associated domains (TADs) -- regions rich in local DNA interactions~\cite{Barbieri2012,NAR,TADs}. 

From a biophysical perspective, a system made up of DNA and DNA-binding proteins exhibits many kinds of interesting and seemingly counter-intuitive behaviour, such as the clustering of proteins in the absence of any attractive interaction between them. This process is driven by the ``bridging-induced attraction''~\cite{Brackley2013}. In conjunction with the specific patterning of binding sites found on a whole human chromosome in vivo, this attraction can drive folding into TADs in the appropriate places on the chromosome~\cite{NAR}. 

In the simple case where there is only a non-specific DNA-protein interaction (i.e., proteins can bind to any point on DNA), bridging-induced clustering can be understood as being due to a thermodynamic feedback loop: binding of bridges to multiple DNA segments causes an increase in local DNA concentration which, in turn, recruits further DNA-binding proteins, and further iterations then sustain the positive feedback. Subsequently, the ensuing clusters coarsen, and eventually phase separate into one macroscopic cluster of DNA-bound bridges in equilibrium with a (diluted) pool of unbound proteins~\cite{Johnson2015,orland}. In the more complex case with specific DNA-binding interactions, clustering is associated with the formation of DNA loops. Looped structures incur an entropic cost which increases superlinearly with the number of loops, and can stop the growth of a cluster beyond a critical size~\cite{NAR,JSTAT,mike,enzo,lagomarsino}. {Such specific binding drives the formation of promoter-enhancer loops~\cite{alberts}; however there are several proteins which interact mainly non-specifically with large regions of the genome, such as histone H1 and other heterochromatin-associated proteins~\cite{alberts}. For this latter class of proteins, the abundance of binding sites in the nucleus would lead to clusters that coarsen progressively. However, this indefinite growth is not observed: we suggest that reversible post-translational protein modifications may be the reason underlying the arrested coarsening.}

Specifically, here we consider a non-equilibrium biochemical reaction which can modify DNA-binding proteins. In our model, these proteins continuously switch between an active, DNA-binding state, and an inactive, non-binding, one. Such a reaction can arise in several scenarios. For instance, a complex of transcription factors and an RNA polymerase might stabilize a promoter-enhancer loop; upon transcription termination, the complex could dissociate and the loop disappear~\cite{alberts,peter}. Alternatively, phosphorylation, or other post-translational modifications of transcription factors~\cite{PTM}, may affect their affinity for chromatin, as might a conformational change in a protein or the reversible addition of a sub-unit to a protein complex, which might be driven by ATP hydrolysis.

In this work we show that introducing this non-equilibrium mechanism strikingly broadens the range of physical behaviour displayed by the chromatin/protein ensemble. In particular, we find that including ``switching'' proteins which interact non-specifically with a chromatin fiber leads to qualitatively and quantitatively different results compared to ``equilibrium'' proteins (which are always in the binding state). Switching (i.e., protein modification) arrests the coarsening triggered by the bridging-induced attraction, and the size of the resulting clusters can be tuned by altering the switching rates. 
Furthermore, we show that if proteins bind both specifically and non-specifically, switching results in the formation of highly-dynamic clusters, which are qualitatively different from those formed by non-switching proteins. In the former case, proteins in the cluster exchange with the soluble pool, whilst the general shape of the cluster persists. These dynamic clusters closely resemble some of the nuclear bodies of eukaryotic cells.
Finally, we consider a simplified model for the formation of TADs in chromosomes, and show that protein switching leaves the location of the domains unaltered, but strongly disfavours long-range inter-TAD interactions. All these findings point to an important and generic role of reversible protein modification in chromatin and nuclear organization. 

\section{Results}

\subsection{Protein switching arrests coarsening of chromatin bridges that bind non-specifically}

We perform Brownian dynamics simulations of a flexible chromatin fiber modeled as a bead-and-spring polymer (thickness 30 nm, persistence length 90 nm) interacting non-specifically with either non-switching or switching proteins. These proteins can bind to the fiber at more than one location (in our case through a Lennard-Jones potential; see \SI{} for details on the force field, and Fig. 1A,B for a schematic). For simplicity, we assume a protein has the same size as the chromatin beads (a realistic assumption as each is likely to be a protein complex). We also assume proteins stochastically switch between binding and non-binding states at an equal rate, $\alpha$. [Relaxation of either of these assumptions does not qualitatively alter results.]

\begin{figure}[t]
\begin{center}
\centerline{\includegraphics[width=0.85\columnwidth]{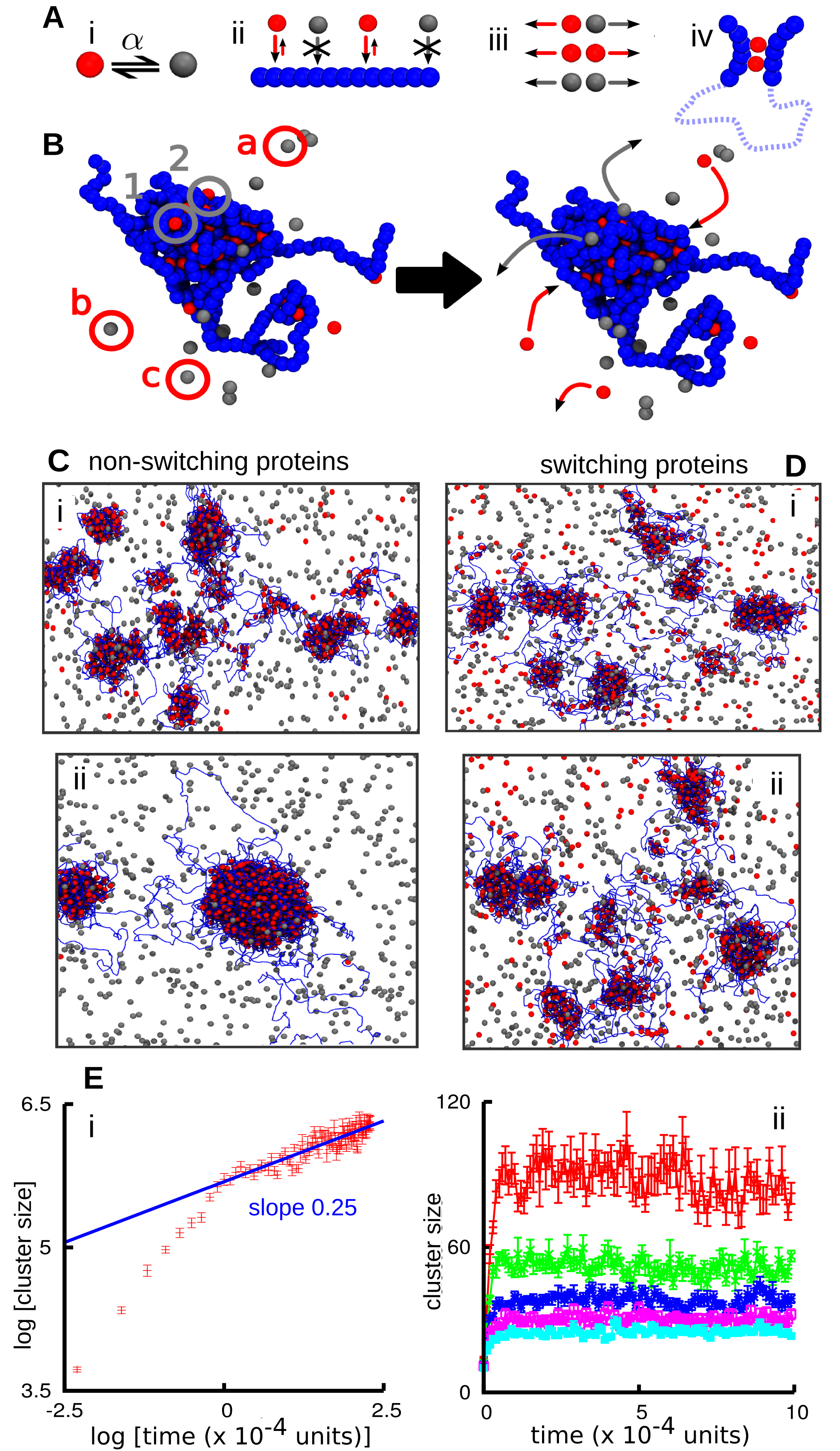}}
\end{center}
\vspace*{-0.5 cm}
\caption{\textbf{Protein switching arrests coarsening.}
In (A-D), active and inactive proteins are colored red and gray respectively; chromatin is represented by strings of blue beads. (A) Schematic of the model (Brownian dynamics simulations). (i) Proteins (lone spheres) switch between red and gray states at rate $\alpha$. (ii) Only proteins in the red state can bind chromatin. (iii) Red and gray beads interact via steric repulsion only. (iv) Proteins can bind to $\ge 2$ sites to create molecular bridges and loops. (B) Snapshots illustrating protein binding/unbinding. Bound active proteins have clustered and compacted chromatin. Bound active proteins 1 and 2 (gray circles) switch and become inactive and dissociate (gray arrows); non-binding proteins a-c in the soluble pool (red circles) become active and sometimes bind to the cluster (red arrows). (C) Snapshots taken (i) $10^4$ and (ii) $2\times 10^4$ simulation units after equilibration. The simulation involved a $5000$-bead fiber (corresponding to 15 Mbp) and $N=4000$ non-switchable proteins, of which half are able to bind. (D) As (C), but for $N=4000$  switchable proteins ($\alpha = 0.0003$ inverse Brownian times). (E) Average cluster size as a function of time. Error bars denote standard deviations of the mean. (i) Non-switching proteins. (ii) Switching proteins; from top to bottom,  $\alpha$ equals $0.0001$, $0.0002$, $0.0003$, $0.0004$ and $0.0005$ inverse Brownian times.}
\label{fig1}
\vspace*{-0.5 cm}
\end{figure}

First, we consider the case of equilibrium, non-switching proteins, where $\alpha=0$ (Fig. 1C). This case was studied in~\cite{Barbieri2012,Brackley2013,Johnson2015,Nicodemi2009}, and it was shown to lead to polymer collapse~\cite{Barbieri2012,Johnson2015,Nicodemi2009} and clustering of bridges~\cite{Brackley2013,Johnson2015} depending on the protein concentration. For the concentrations used here, clusters coarsen and grow at the expense of smaller aggregates. 
During the early stages, this resembles the Ostwald ripening characteristic of liquid-gas phase separation; later on, we also observe coalescence of smaller clusters into larger ones (Movie S1). The average cluster size -- measured as the number of bound proteins per cluster -- increases with time with no sign of saturation until all clusters merge into one (Fig. 1Cii, and Movie S1). For early times, cluster size (which is also proportional to its volume) increases approximately linearly with time, as would be expected for Ostwald ripening in density-conserving model B~\cite{Chaikin2000}. For later times, cluster growth is much slower, with a sublinear exponent (close to 0.25 for our parameters, see Fig. 1Ei). This slowing is due to the underlying polymer dynamics -- as in blob formation during the collapse of a homopolymer, which is also slower than simple model B kinetics~\cite{polymercollapsekinetics}.

The dynamics with protein switching is remarkably different: now, coarsening is completely arrested, and the system achieves a micro-phase separated state in which clusters have a well-defined average size (Fig. 1D; Movie S2). This size decreases with $\alpha$ (Fig. 1Eii). The arrested phase separation can be understood intuitively as follows. On the one hand, thermodynamics dictates that the system should try to minimize interfaces, and this leads to coarsening, initially via Ostwald ripening, giving the growth laws in Figure 1Ei. On the other hand, non-equilibrium protein switching is a Poisson process, so active proteins switch off at a constant rate $\alpha$, and leave the cluster. [This is not the case for equilibrium proteins which can only unbind thermodynamically; not only is the unbinding rate slower, but such proteins are also highly likely to rebind to a nearby site before ever leaving a cluster]. Then, active proteins only have a timescale of the order $\alpha^{-1}$ in which to form a cluster, before a significant proportion of proteins in that cluster inactivate. Hence, phase separation is arrested. 

\subsection{A mean field theory quantitatively explains the arrest of coarsening, and predicts average cluster size}
To understand more quantitatively how a non-equilibrium biochemical reaction arrests coarsening, we consider a simplified mean field theory which follows the time evolution of the chromatin density {$\rho(\bm{x},t)$}, and of the active protein density $\Phi(\bm{x},t)$. Our equations describe the binding of the proteins to the chromatin together with the diffusion of all components, and they read as follows,
\begin{eqnarray}\label{meanfield1}
\dot \rho &=& M_{\rho}\nabla^2\left[a_1 \rho - k \nabla^2 \rho - \chi \Phi + g \rho^3\right]
\\ \nonumber 
\dot \Phi &=& M_{\Phi} \nabla^2 \left[a_2 \Phi - \chi \rho \right] - \alpha (\Phi-\Phi_0).
\end{eqnarray}
These equations can be formally derived starting from a suitable underlying free energy density, and adding protein modification as a reaction term -- the details are discussed in the \SI{}. In Eqs.~(\ref{meanfield1}), $M_{\rho}$ and $M_{\Phi}$ are the chromatin and protein mobility respectively, so that $M_{\rho}a_1\equiv D_1$ and $M_{\Phi}a_2\equiv D_2$ represent effective diffusion coefficients, while $\chi$ is the coefficient describing bridging between active proteins and chromatin. Further, $g$ captures steric repulsion in the chromatin fiber, $\kappa$ accounts for effective surface tension effects, and finally the last term in the equation for $\Phi$ models the biochemical reaction, where proteins switch from binding to non-binding, and vice versa, at a rate $\alpha$. For $\alpha=0$, Eqs.~(\ref{meanfield1}) ensure conservation of the global density of both chromatin and proteins -- in other words, this is an example of generalized model B dynamics~\cite{Berry2015,Chaikin2000}.

To identify the key parameters in our system, we now choose dimensionless time and space units $t_u=1/\alpha$ and $x_u=\sqrt{D_2/\alpha}$ and redefine $\Phi$ as $\Phi(M_\rho \chi/D_2)$. In these units, our equations become 
\begin{eqnarray}
\dot \rho &=& \mathcal{D}_0 \nabla^2 \rho - A\nabla^4 \rho - \nabla^2 \phi + G \nabla^2 \rho^3 \label{rhod} \\
\dot \Phi &=& \nabla^2 \Phi - X \nabla^2 \rho - (\Phi-\Phi_0), \label{phid} 
\end{eqnarray}
so that the whole parameter space is spanned by the four dimensionless numbers $X=(\chi^2 M_\rho M_\Phi/D_2^2)$; $\mathcal{D}_0=(D_1/D_2)$; $A=\alpha k M_\rho/(D_2^2)$ and $G=gM_\rho/D_2$.

One solution of Eqs.~(\ref{rhod}-\ref{phid}) is given by the uniform phase $(\rho,\Phi)=(\rho_0,\Phi_0)$, which is stable in the absence of bridging ($\chi=0$). To see how the interplay of bridging and biochemical switching can create patterns, we performed a linear stability analysis of this uniform state (Fig. 2, detailed in the \SI{}). The result is that small perturbations of the uniform phase grow if $X>X_{\rm c}=(\sqrt{A}+\sqrt{\mathcal{D}})^2$ where $\mathcal{D}=\mathcal{D}_0+3G\rho_0^2$. This instability criterion translates in physical units to $\chi>\sqrt{k\alpha/M_\Phi}+\sqrt{(D_2/M_\Phi)[D_1/M_\rho +3g \rho_0^2]}$. Thus, the instability is driven by bridging, whereas diffusion of chromatin and proteins, excluded volume, as well as protein modification, all tend to stabilize uniform chromatin-protein distributions. Importantly, this bridging-induced instability also works at very low protein concentration.

Calculating the wavenumber at the onset of instability (see \SI{}) unveils the remarkable role played by the biochemical reaction for structure formation. Specifically, we find $q_c=\left(\mathcal{D}/A\right)^{(1/4)}$ for the dimensionless onset wavenumber, translating in physical units to the following typical length scale,
\1 \label{lengthscale}
L_c = 2\pi  \left[ \4{D_1 D_2 + 3M_\rho g\rho_0^2 D_2}{\alpha k M_\rho} \right ]^{1/4}.\2
Hence, in contrast to models without protein modification, the present system exhibits a short wavelength instability (Fig.~2), which turns into a long wavelength instability only in the limit $\alpha \rightarrow 0$ (which would lead to $L_c \rightarrow \infty$, dotted black line in Fig.~2). Our linear stability analysis therefore suggests that the presence of the biochemical reaction has {\it qualitative} consequences for the clustering in the system, in that it leads to self-limiting cluster sizes, or put differently, to micro-phase separation rather than to macro-phase separation -- in full agreement with the simulations shown in Fig.~1. 

To further confirm that within our mean field theory clusters cannot coarsen indefinitely, we also performed a weakly non-linear expansion, through which we found that the amplitude of the chromatin density fluctuations close to the uniform state obey the ``real Ginzburg Landau equation'', which is associated with formation of stationary patterns of well-defined self-limiting size~\cite{Cross1993} (see \SI{}). Finally, Eq.~(\ref{lengthscale}) also predicts that, at least close to the onset of clustering, the average number of proteins in any aggregate should scale as $L_c^3 \sim \alpha^{-3/4}$, in good agreement with results from our simulations (Fig. S1).

\begin{figure}[t]
\centering
\includegraphics[width=0.85\columnwidth]{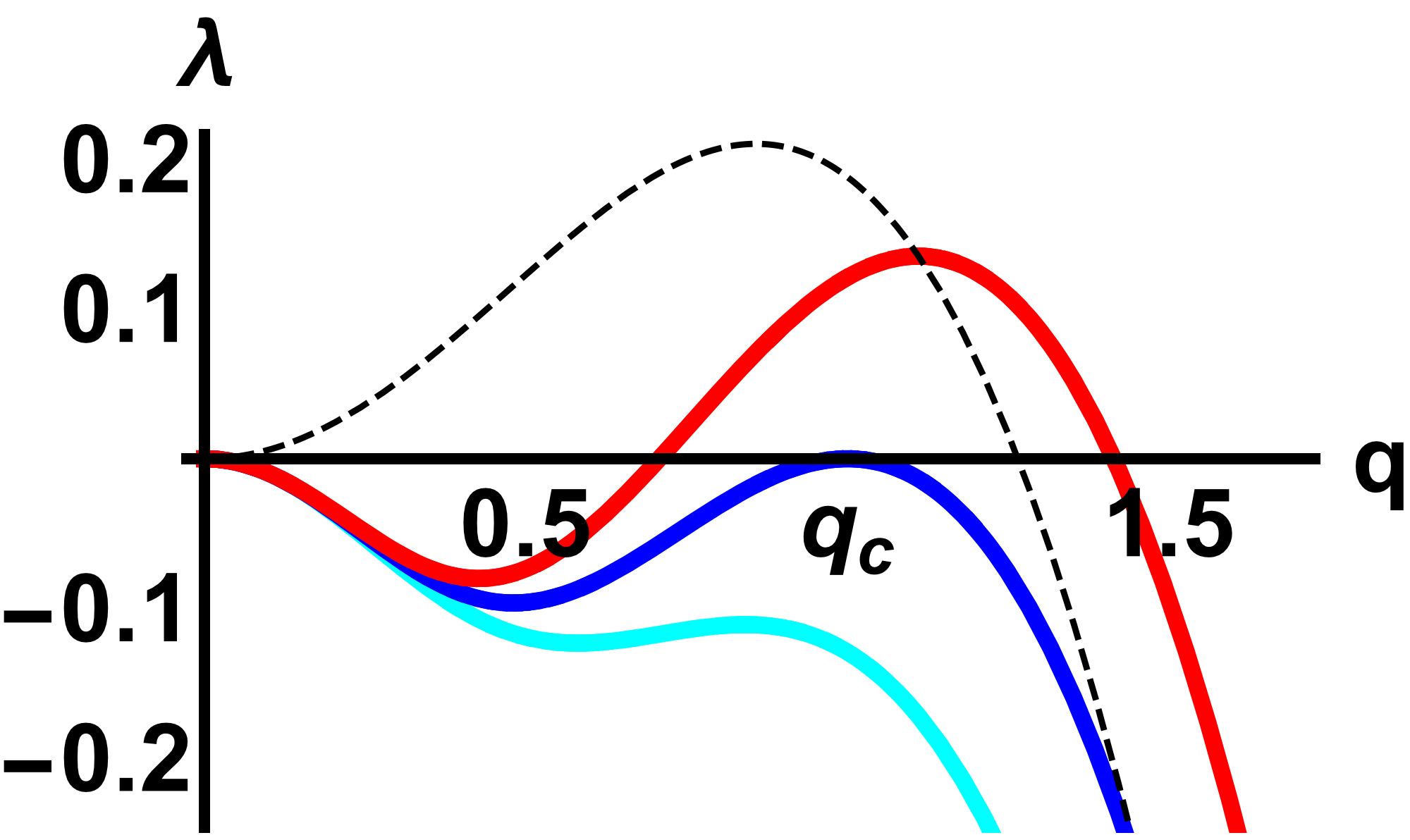}
\caption{\textbf{Mean field theory predicts arrested coarsening with protein modification.} Dispersion relation, showing the growth rate, $\lambda$, as a function of the magnitude of the wavevector, $q$, for fluctuations around the uniform solution of Eqs.~(\ref{rhod}), for $\mathcal{D}=A=1$, and $X=3.5$ (cyan), corresponding to linear stability of the uniform phase, $X=X_{\rm c}=4.0$ (blue), marking the onset of instability, and $X=4.5$ (red), revealing the growth of clusters with a characteristic length scale. The dotted black line shows a typical dispersion relation in the absence of protein modification, which leads to a long wavelength instability.}
\label{fig2}
\end{figure}

\begin{figure*}
\centerline{\includegraphics[width=0.9\textwidth]{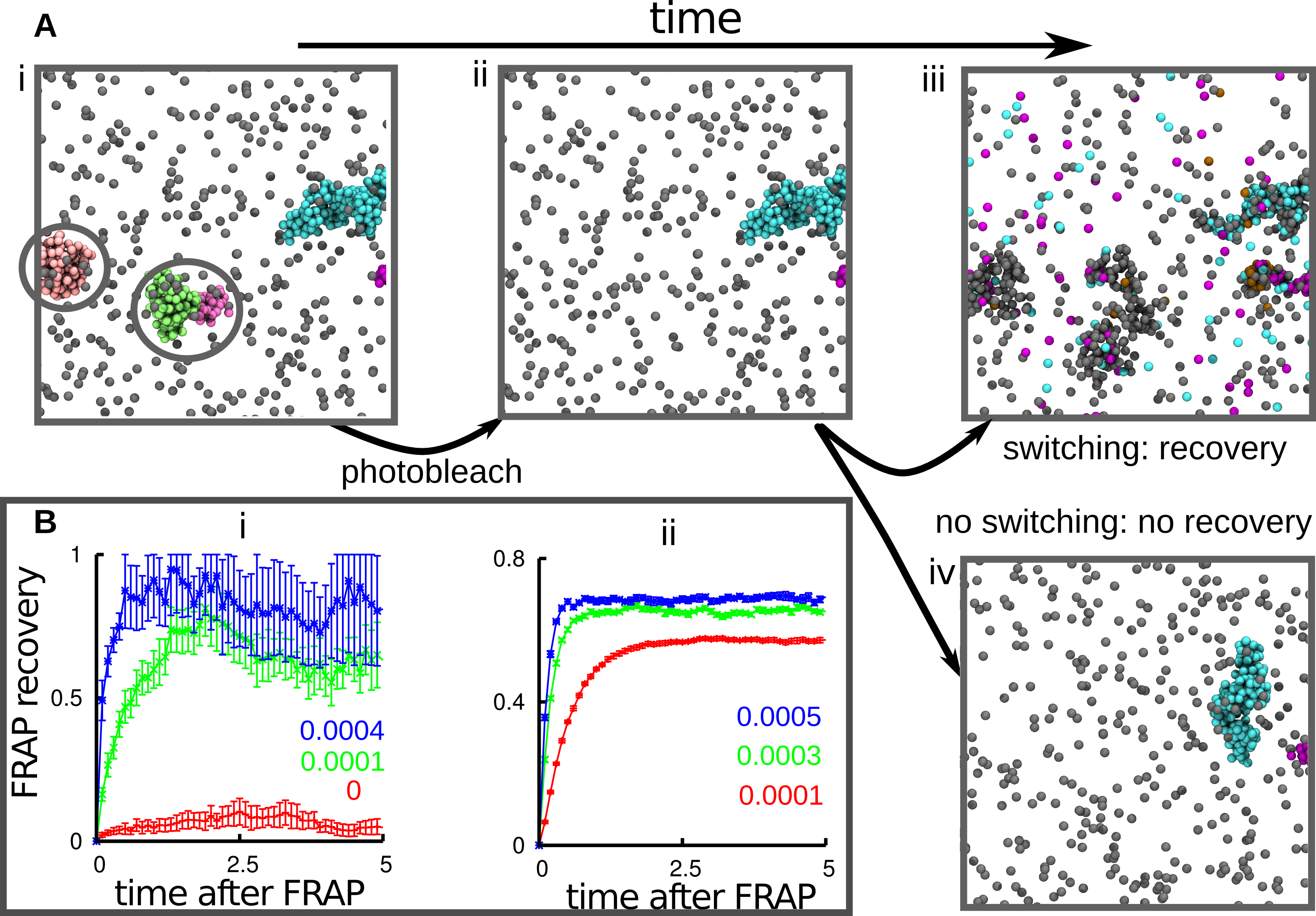}}
\caption{\textbf{{\it In silico} FRAP (Brownian dynamics simulations).} (A) Snapshots taken $10^4$ (i,ii) or $2\times 10^4$ (iii,iv) after equilibration, during an {\it in silico} FRAP experiment (only proteins -- and not chromatin beads -- are shown for clarity). (i) The simulation begins with $N=2000$ equilibrium proteins, half of which are able to bind the chromatin fiber, both specifically (interaction strength $15$ $k_BT$, cut-off $1.8 \sigma$) to every 20-th bead in the polymer, and non-specifically (interaction strength $4$ $k_BT$, cut-off $1.8 \sigma$) to any other bead. After $10^4$ time units, a structure with multiple clusters forms. The snapshot shows only a portion of this, for clarity; $5$ clusters of bound proteins have developed (unbound proteins are colored gray, and bound ones in the $5$ clusters a different color). Circled areas will be ``photo-bleached''. (ii) Photo-bleaching involves making bound proteins invisible (the bleached proteins are still present in the simulation). (iii) If proteins can switch, clusters reappear in the same general place (as new proteins replace their ``bleached'' counterparts). (iv) If proteins cannot switch (i.e., $\alpha = 0$), clusters do not recover (as their protein constituents do not recycle). (B) FRAP recovery. Error bars give SD of mean, and time is given in multiples of $10^4$ simulation units; the values of $\alpha$, in units of inverse Brownian times, are as indicated in each panel. Only the post-bleaching signal is shown (the pre-bleaching value would be constant and equal to $1$ in these units). (i) Number of unbleached proteins in the bleached volume (a sphere of $50\sigma$) as a function of time, after bleaching. The signal is normalized with respect to the number of proteins initially in the bleached volume. (ii) Number of unbleached proteins in clusters as a function of time after bleaching, after all proteins in clusters at a given instant are bleached. The signal is normalized with respect to the proteins in clusters at the time of bleaching.}
\label{fig3}
\end{figure*}

\subsection{Switching proteins with specific binding self-assemble into recycling nuclear bodies}

The model considered in Figure 1 assumes proteins bind non-specifically. While this is a good approximation for generic heterochromatin-binding proteins in silenced regions, most transcription factors bind to active regions specifically and to most other DNA non-specifically~\cite{kafri}. 
Therefore, we consider proteins binding with high affinity to every $20$th bead (i.e., every 60 kbp), and with low affinity to all others. [Similar results are expected for different patterns of binding sites~\cite{NAR,Brackley2013}.] Now bound proteins self-assemble into clusters of self-limiting size even when $\alpha=0$ (Fig.~3; Movie S3). In other words, coarsening is always arrested. As suggested previously ~\cite{Brackley2013,JSTAT,enzo}, specific binding creates loops and loop clustering is associated with entropic costs that scale super-linearly with loop number, and this limits cluster growth~\cite{JSTAT}.

Although coarsening is arrested whatever the value of $\alpha$, there is a major difference between the dynamics of the equilibrium and switching proteins. Without switching, proteins can only unbind thermodynamically, which requires a long time: as a result, proteins rarely exchange between clusters (Fig. 3, Movie S3). With switching, there is a constant turnover of proteins within the clusters, which recycle all their components over a time $\sim\alpha^{-1}$ (Fig. 3, Movie S4). Reducing the strength of the specific interactions can also lead to protein turnover (Fig. S2), but this requires fine tuning of the parameters so as to simultaneously ensure stable binding and the recycling of proteins in clusters. In contrast, protein modification naturally leads to such recycling for any values of specific and non-specific binding affinity.

To quantitatively characterize the dynamics of turnover within clusters, we perform a simulated fluorescence-recovery-after-photobleaching (FRAP) experiment~\cite{FRAP}. In such an experiment some of the clusters are photobleached at a given time, and recovery of fluorescence is then monitored (Fig. 3). The ``fluorescence'' signal (proportional to the number of non-photobleached active proteins in the clusters) recovers quickly in the $\alpha\ne 0$ case (Figs. 3Aiii, 3B), but not in the $\alpha=0$ case (Fig. 3Aiv, 3Bi), at least for large values of the specific interaction strength. The dynamics of recovery can be measured using the number of unbleached proteins in the photobleached volume (Fig. 3Bi); this is proportional to the fluorescence intensity measured in a standard FRAP experiment. Alternatively, the number of unbleached proteins in clusters can be used (Fig. 3Bii). Both approaches give similar recovery timescales, and confirm that protein modification is required to create clusters in which proteins can recycle. 

The clusters found in Figure 3 typically contain $\sim 20-100$ proteins that recycle (Fig. S3A). Cluster size depends on both protein concentration and interaction energy (e.g., in Fig. S3B there are only $\sim 5-10$ proteins per cluster). Therefore, this mechanism can produce clusters with a wide range of sizes. Note that nuclear bodies range from large nucleoli (up to several $\mu$m), through Cajal, polycomb, and promyelocytic leukemia bodies ($\sim 1$ $\mu$m)~\cite{Sleeman2014,Mao2011,Pirrotta2012}, to transcription factories containing $\sim 10$ active transcription complexes ($\sim 100$ nm)~\cite{peter,Papantonis2013,Marenduzzo2006}. Importantly, like most nuclear bodies, our clusters also retain a ``memory'' of their shape. Thus, in Figure 3A, when most of the components of the pink cluster on the left have turned over, the general shape of the cluster persists (see also Movies S4 and S5, and Fig. S4). This is because the chromatin scaffold associated with the protein clusters (i.e., the sites of specific binding) retains a general 3D structure that does not change much over time (Fig. S5). Taken together, these results strongly support the conjecture that nuclear bodies emerge from the aggregation of bound switching proteins, and that switching both arrests phase separation and ensures that bound proteins continually exchange with the soluble pool.
	
Notably, the nuclear bodies which our clusters resemble generally show FRAP recovery times in the range of tens of minutes~\cite{FRAPCB,FRAPPcG,Kimura2002}. These time scales are too slow to be accounted for by diffusion, and too fast to be compatible with the thermodynamic unbinding of tightly bound proteins (see \SI{}). Our results suggest an attractive alternative explanation: that the recovery time, over which nuclear bodies recycle their proteins, is instead linked to protein modification, and it is simply proportional to $\alpha^{-1}$. Typical rates of post-translational protein modification can be of the order of several seconds (and will be slower within nuclear bodies due to macromolecular crowding), and transcription termination occurs minutes after initiation. In light of this, our simulations would predict recovery time scales of the order of $\alpha^{-1}$, or minutes, broadly in agreement with those measured experimentally~\cite{FRAPCB,FRAPPcG,Kimura2002}. Further to this, there is biological evidence that protein modifications can take place within nuclear bodies~\cite{FRAPCB}. For instance, enzymes performing post-translational modifications are found in Cajal bodies~\cite{FRAPCB}, and phosphorylation or ubiquination of the BMI1 subunit of the PcG PRC1 complex are important factors which determine the kinetics of exchange in polycomb bodies~\cite{FRAPPcG}. 

\subsection{Protein switching preserves TAD structure, while suppressing long-range interactions}
 
Clustering of bridging proteins can lead to the formation of chromatin domains~\cite{Barbieri2012,NAR,Davide,Brackley2013} resembling TADs found in Hi-C data ~\cite{TADs}. It is therefore of interest to ask how switching affects TAD structure and dynamics. Here, we return to a toy model first considered elsewhere: the fiber has a regular pattern of binding and non-binding regions (Fig.~4A), and each binding region spontaneously and reproducibly assembles into a TAD which is flanked by a disordered non-binding region~\cite{NAR}. 
The regular interspersion of non-binding segments in Figure 4A fixes the locations of TAD boundaries; consequently, clusters form (Fig. 4B,C) at reproducible positions along the fiber, and this -- in turn -- yields TADs seen in averaged contact maps (Fig. 4D). Such patterns resemble those seen in Hi-C data obtained from cell populations.

Variations in $\alpha$ have several effects (Fig. 4). First, the configurations found at steady state are qualitatively different. Although cluster growth is limited for both $\alpha=0$ and $\alpha > 0$, the (recycling) clusters formed by switching proteins are much smaller (Figs. 4B, C, and Movie S6). Second (and notwithstanding this qualitative difference), the contact maps close to the diagonal are remarkably similar (Fig. 4D; compare patterns on each side immediately next to the diagonal); this indicates that local TAD structure is largely unperturbed by switching. However, for $\alpha \ne 0$ non-local (i.e., distant) contacts are strikingly suppressed (Fig. 4D, compare patterns on each side far from the diagonal), and higher order folding of one TAD onto another is suppressed. This was demonstrated directly as follows. First, TADs were generated in the presence of non-switching proteins, and then switching turned on; the fraction of non-local contacts falls (Figs 4E, S6). 

This behaviour can be explained as follows. First, the time scale for the formation of TADs is comparable to (or smaller than) that of protein recycling within  a TAD (see \SI{} for an estimate of such time scales). Computer simulations of TAD formation in {\it Drosophila} and human chromosomes also suggest that the local structure is formed very rapidly (at most, in minutes)~\cite{NAR,Davide}. Therefore it is plausible that local TAD folding is fast enough not to be perturbed much by protein modification. Second, when a particular protein switches from binding to non-binding, a contact is lost, and it is likely that local ones can reform faster than non-local ones. Furthermore, switching provides a non-equilibrium mechanism allowing faster large-scale rearrangements, and so a more effective trimming of entropically unfavourable long-ranged interactions. In other words, active post-translational modification tilts the balance in favour of local intra-TAD contacts at the expenses of inter-TAD ones. This is consistent with the sharp decay beyond the Mbp scale seen in Hi-C data~\cite{TADs,LiebAiden}.   

\begin{figure}[t]
\centerline{\includegraphics[width=1.\columnwidth]{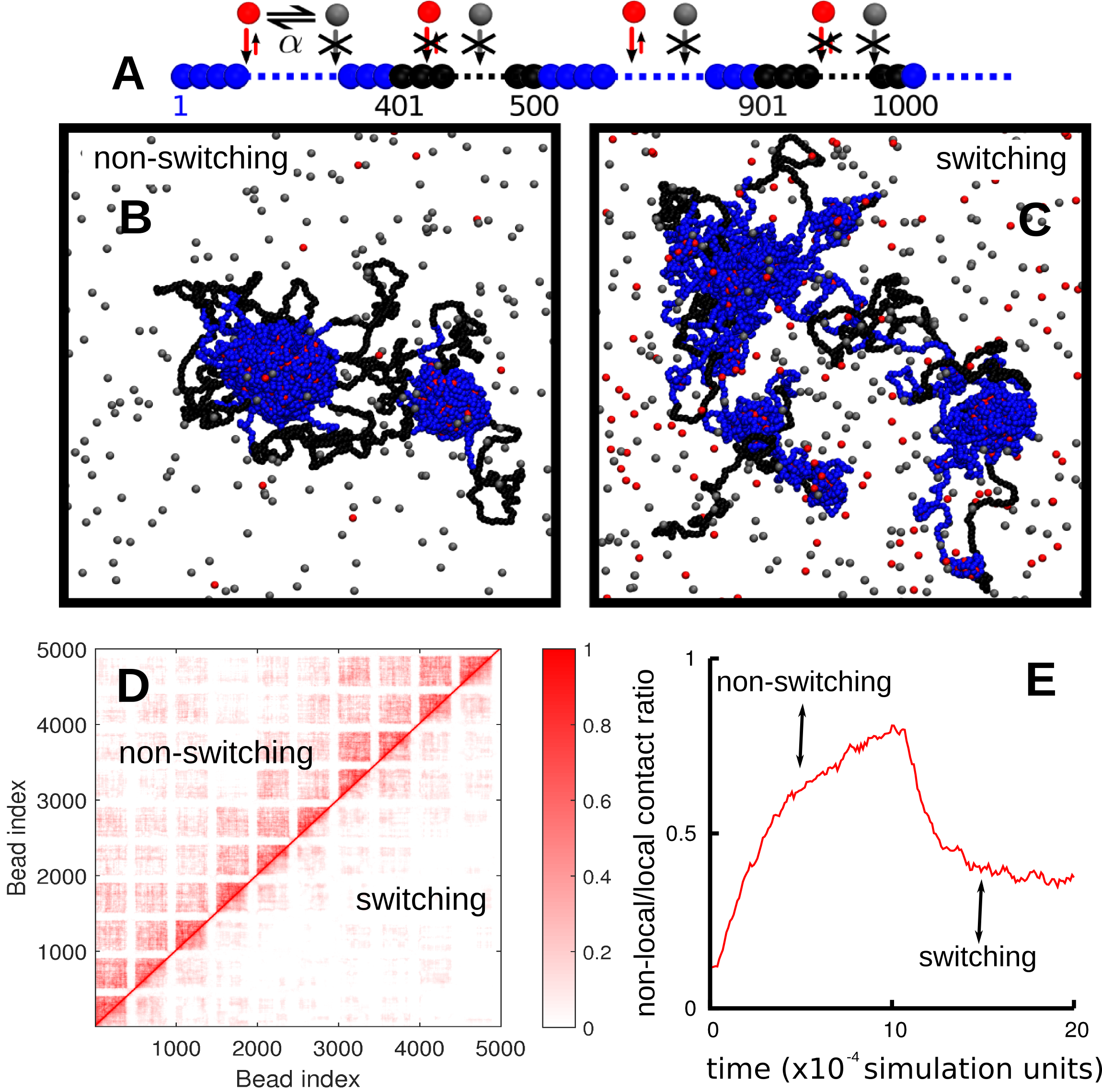}}
\caption{\textbf{Switching promotes intra-TAD contacts and suppresses inter-TAD ones.} (A) Overview. Simulations involved $N=2000$ non-switching ($\alpha = 0$) or $N=2000$ switching proteins ($\alpha= 0.0001$ inverse Brownian times); for $\alpha=0$ half of the proteins are binding. In both cases, interaction  energy and cut-off were $4$ $k_B T$ and $1.8\sigma$. The fiber (length 15 Mbp) consisted of regularly-interspersed segments containing runs of binding (blue) and non-binding (black) beads (segment sizes 900 and 300 kbp, respectively). (B,C) Snapshots taken after $10^5$ simulation units. Non-binding and binding proteins are colored gray and red, respectively.  (D) Contact maps (averages from 10 simulations) for non-switching (top-left triangle), and switching proteins  (lower-right triangle). The scale (right) indicates contact frequencies. (E) The evolution of the ratio of non-local contacts over time. A local (non-local) contact is one between beads separated by less (more) than $1.2$ Mbp along the fiber. Here the simulation was run for $10^5$ simulation units with non-switching proteins; switching was then turned on ($\alpha=0.0001$ inverse Brownian times) and the simulation was run for a further $10^5$ simulation units.}
\label{fig4}
\end{figure}

\vspace*{-0.5 cm}
\section{Conclusions}

To conclude, we have shown that active post-translational protein modication (e.g., phosphorylation, methylation, acetylation~\cite{PTM}, or any other non-equilibrium reaction where a protein switches between binding and non-binding states) has a profound and generic effect on the behaviour of a chromatin-protein mixture. The interplay between protein bridging and protein modification is therefore an important principle underlying nuclear organization within eukaryotes.

First, it was previously shown that non-switching proteins able to bind non-specifically to chromatin to form molecular bridges assemble into clusters which have a natural tendency to coarsen~\cite{Brackley2013,Johnson2015}. Here we show that switching changes the behaviour; cluster growth is self-limiting (Fig. 1) -- a phenomenon which be understood via a simple mean field theory (Fig. 2). This theory provides an example of arrested phase separation, and it can explain why nuclear bodies do not progressively enlarge~\cite{nuclearbodies,Sleeman2014,Mao2011,Pirrotta2012}, and why neighbouring clumps of heterochromatin -- whether detected using classical staining and microscopy, or through inspection of Hi-C contact maps~\cite{TADs,Sexton2012} -- rarely merge into one super-domain.

Second, when equilibrium (i.e., non-switching) proteins bind specifically to cognates sites on the chromatin fiber, they also cluster; however, specific binding is known to arrest the coarsening~\cite{NAR,Brackley2013}.  But in contrast to what is seen in photobleaching experiments~\cite{FRAPCB,FRAPPcG}, bound proteins in the ensuing clusters exchange little with the soluble pool. Moreover, the time scales seen in such bleaching experiments are too slow to be accounted for by diffusion, and too fast to be compatible with the thermodynamic unbinding of tightly-bound proteins. Protein modification provides a neat solution to this paradox: dynamic clusters naturally emerge during simulations, with constituent proteins recycling on a time-scale proportional to the inverse switching rate, $\alpha^{-1}$ (Fig. 3). Importantly, when clusters in simulations are ``photobleached'', they behave like nuclear bodies seen in vivo -- they retain a ``memory'' of their shape, despite the continual exchange with the soluble pool.

Finally, switching affects large-scale chromatin organization. Bridging-induced clusters are associated with the formation of chromatin domains, reminiscent of the TADs observed in Hi-C data~\cite{TADs}. Using a fiber patterned in such a way that it spontaneously folds into TADs, we find that switching has little effect on local TAD organization, but strongly suppresses inter-TAD interactions; local contacts are favored over non-local ones (Fig. 4). We expect that similar trends should be observed in more complex models for bridging-induced chromosome organization, such as those in Refs.~\cite{Barbieri2012,NAR,Brackley2013}.

While here we focus on a flexible chromatin fiber, we expect that near-identical results should be reached with a semi-flexible one~\cite{Brackley2013,orland}; then, our conclusions should also apply to bacterial DNA. We also expect similar results to be obtained with more complex pathways between active and inactive states (e.g., modeling the cyclic flooding of proteins into nuclei, or their cyclic synthesis/degradation) and it would be of interest to investigate these scenarios.
	
In summary, we demonstrated how non-equilibrium processes involving ephemeral protein states can provide a simple way of understanding how dynamic nuclear bodies of self-limiting size might form, and how chromosomal domains at the larger scale might be organized. 

This work was funded by ERC (Consolidator Grant 648050, THREEDCELLPHYSICS), and by a Marie Skodowska-Curie Intra European Fellowship (G. A. No. 654908). 
\vspace*{-0.2 cm}

\bibliographystyle{pnas}

\begin{thebibliography}{99}
\bibitem{understandingDNA} Calladine, C.~R., Drew, H.~R., Luisi, B.~F., Travers, A.~A. (2004). {\it Understanding DNA}, 3rd Edition, Elsevier Academic Press, London.
\bibitem{alberts} Alberts, B. {\it et al.} (2002). {\it Molecular biology of the
cell}, Garland Science, New York.
\bibitem{peter} Cook, P.~R. (2001). {\it Principles of Nuclear Structure and Function}, Wiley-Liss (New York) (2001).
\bibitem{HNS} Dame, R.~T., M.~C. Noom, M.~C., Wuite,  G.~J. (2006). 
Bacterial chromatin organization by H-NS protein unravelled using dual DNA manipulation. {\it Nature} {\bf 444}, 387-390. 
\bibitem{Simonis2006}
Simonis, M. {\it et al.} 
(2006) Nuclear organization of active and inactive chromatin domains uncovered by chromosome conformation capture-on-chip (4C). {\it Nat. Genet.} {\bf 38}, 1348-1354.
\bibitem{Barbieri2012}
Barbieri, M. {\it et al.} 
(2012). Complexity of chromatin folding is captured by the strings and binders switch model. {\it Proc. Natl. Acad. Sci. USA} {\bf 109}, 16173-16178.
\bibitem{GenomeBiology} Brackley, C.~A. {\it et al.} 
(2016). Predicting the three-dimensional folding of cis-regulatory regions in mammalian genomes using bioinformatic data and polymer models. {\it Gen. Biol.} {\bf 17}, 59 (2016).
\bibitem{NAR} Brackley, C.~A., Johnson, J., Kelly, S., Cook, P.~R., Marenduzzo, D. (2016). Simulated binding of transcription factors to active and inactive regions folds human chromosomes into loops, rosettes and topological domains. {\it Nucl. Acids Res.} {\bf 44}, 3503-3512.
\bibitem{HP1} Kilic, S., Bachmann, A.~L., Bryan, L.~C.,  Fierz, B. (2015).
Multivalency governs HP1$\alpha$ association dynamics with the silent chromatin state. {\it Nat. Commun.} {\bf 6}, 7313 (2015).
\bibitem{H1} Harshman, S.~W., Young, N.~L., Parthun, M.~R., Freitas, M.~A. (2013). H1 histones: current perspectives and challenges. {\it Nucl. Acids Res.} {\bf 41}, 9593-9609 (2013). 
\bibitem{Ajaz} Wani, A.~H. {\it et al.} 
(2016). Chromatin topology is coupled to Polycomb group protein subnuclear organization.  {\it Nat. Comm.} {\bf 7}, 10291. 
\bibitem{Davide} Michieletto, D., Marenduzzo, D., Wani, A.~H. (2016) Chromosome-wide simulations uncover folding pathway and 3D organization of interphase chromosomes. arXiv:1604.03041.
\bibitem{nuclearbodies} Zhu, L., Brangwynne, C.~P.  (2015). Nuclear bodies: the emerging biophysics of nucleoplasmic phases. {\it Curr. Opin. Cell Biol.} {\bf 34}, 23-30.  
\bibitem{Sleeman2014} Sleeman, J.E., Trinkle-Mulcahy, L. (2014). Nuclear bodies: new insights into assembly/dynamics and disease relevance. {\it Curr. Opin. Cell Biol.} {\bf 28}, 76-83.
\bibitem{Mao2011} Mao, Y.~S., Zhang, B., Spector, D.~L. (2011) Biogenesis and function of nuclear bodies. {\it Trends Genet.} {\bf 27}, 295-306. 
\bibitem{Pirrotta2012} Pirrotta, V., Li, H.~B. (2012) A view of nuclear polycomb bodies. {\it Curr. Opin. Gen. Devel.} {\bf 22}, 101-109.  
\bibitem{Berry2015} Berry, J., Weber, S.~C., Vaidya, N., Haataja, M., Brangwynne, C.~P. (2015). RNA transcription modulates phase transition-driven nuclear body assembly. 
\bibitem{Brangwynne2015} Brangwynne C.~P., Tompa P., Pappu R.~V., (2015) Polymer physics of intracellular phase transitions. {\it Nat Phys.} {\bf 11}, 11, 899-904. 
\bibitem{TADs} Dixon, J.~R., Selvaraj, S., Yue, F., Kim, A., Li, Y., Shen, Y., Hu, H., Liu, J.~S., Ren, B. (2012).
Topological domains in mammalian genomes identified by analysis of chromatin interactions. {\it Nature} {\bf 485}, (2012) 376-380.
\bibitem{Brackley2013} 
Brackley, C.~A, Taylor, S., Papantonis, A., Cook, P.~R., Marenduzzo, D. (2013). Nonspecific bridging-induced attraction drives clustering of DNA-binding proteins and genome organization. {\it Proc. Natl. Acad. Sci. USA} {\bf 110}, E3605-3611. 
\bibitem{Johnson2015} Johnson, J., Brackley, C.~A., Cook, P.~R., Marenduzzo, D. (2015). A simple model for DNA bridging proteins and bacterial or human genomes: bridging-induced attraction and genome compaction. {\it J. Phys. Cond. Matt.} {\bf 27}, 064119. 
\bibitem{orland} Le Treut, G., Kepes, F., Orland, H. (2016).
Phase Behavior of DNA in the Presence of DNA-Binding Proteins
 {\it Biophys. J.} {\bf 110}, 51-62 (2016). 
\bibitem{JSTAT} Marenduzzo, D., Orlandini, E. (2009). Topological and entropic repulsion in biopolymers. {\it JSTAT}, L09002. 
\bibitem{mike} Cates, M.~E., Witten, T.~A. (1986). Chain conformation and solubility of associating polymers.  {\it Macromolecules} {\bf 19}, 732-739.
\bibitem{enzo} Orlandini, E., Garel, T. (1998). Collapse transitions of a periodic hydrophilic hydrophobic chain, {\it Eur. Phys. J. B} {\bf 6}, 101-110.
\bibitem{lagomarsino} Scolari, V.~F., Lagomarsino, M.~C. (2015).
Combined collapse by bridging and self-adhesion in a prototypical polymer model inspired by the bacterial nucleoid. {\it Soft Matter} {\bf 11}, 1677-1687.
\bibitem{PTM} Tootle, T.~L., Rebay, I. (2005). Post-translational modifications influence transcription factor activity: a view from the ETS superfamily. {\it Bioessays} {\bf 27}, 285-298 (2005).
\bibitem{Nicodemi2009} Nicodemi, M., Prisco, A. (2009).
Thermodynamic Pathways to Genome Spatial Organization in the Cell Nucleus.
{\it Biophys. J.} {\bf 96}, 2168-2177.
\bibitem{Chaikin2000} Chaikin, P.~M., Lubensky, T.~C. (2000). 
{\it Principles of Condensed Matter Physics}, Cambridge University Press. 
\bibitem{polymercollapsekinetics} Byrne, A., Kiernan, P., Green, D., Dawson, K.~A. (1995). Kinetics of homopolymer collapse. {\it J. Chem. Phys.} {\bf 1012}, 573-577 (1995).
\bibitem{Cross1993} Cross, M.~C., Hohenberg, P.~C. Pattern formation outside of equilibrium. {\it Rev. Mod. Phys} {\bf 65}, 851-1112 (1993).
\bibitem{kafri} Sheinman, M., Benichou, O., Kafri, Y., Voituriez, R. (2012).
Classes of fast and specific search mechanisms for proteins on DNA.
{\it Rep. Progr. Phys.} {\bf 7}, 026601. 
\bibitem{FRAP} Mueller, F., Mazza, D., Stasevich, T.~J., McNally, J.~G. (2010).FRAP and kinetic modeling in the analysis of nuclear protein dynamics: what do we really know? {\it Curr Opin Cell Biol.} {\bf 22}, 403-411. 
\bibitem{Papantonis2013} Papantonis, A., Cook, P.R. (2013). Transcription factories; genome organization and gene regulation. {\it Chem. Rev.} {\bf 113}, 8683-8705.
\bibitem{Marenduzzo2006}
Marenduzzo, D., Micheletti, C., Cook, P.~R. (2006). Entropy-driven genome organization. {\it Biophys. J.} {\bf 90}, 3712-3721.
\bibitem{FRAPCB} Handwerger, K.~E., Murphy, C., Gall, J.~G. (2003)
Steady-state dynamics of Cajal body components in the Xenopus germinal vesicle.
{\it J. Cell. Biol.} {\bf 160}, 495-504.
\bibitem{FRAPPcG}
Hern\'andez-Mu\~noz I, Taghavi P, Kuijl C, Neefjes J, van Lohuizen M. (2005)
Association of BMI1 with polycomb bodies is dynamic and requires PRC2/EZH2 and the maintenance DNA methyltransferase DNMT1.
{\it Mol Cell Biol.} {\bf 25} 11047-58.
\bibitem{Kimura2002} Kimura, H., Sugaya, K., Cook, P.~R. (2002)
The transcription cycle of RNA polymerase II in living cells. {\it J. Cell Biol.} {\bf 159}, 777-782. 
\bibitem{LiebAiden} Sanborn, A.~L. {\it et al.} (2015). Chromatin extrusion explains key features of loop and domain formation in wild-type and engineered genomes. {\it Proc. Natl. Acad. Sci. USA} {\bf 112}, E6456-E6465
\bibitem{Sexton2012} Sexton, T. {\it et al.} 
(2012). Three-dimensional folding and functional organization principles of the Drosophila genome. {\it Cell} {\bf 148}, 458-472.
\end{thebibliography}

\begin{thebibliography}{99}	
\bibitem{Kremer1990} Kremer K., Grest G.~S. (1990) Dynamics of entangled linear polymer melts: A molecular-dynamics simulation. {\it J. Chem. Phys.} 92(8):5057.
\bibitem{Chaikin2000}
P.M. Chaikin, P.~M., Lubensky, T.~C. (2000). {\it Principles of condensed matter physics}, Cambrdige University Press, Cambridge.
\bibitem{Cross1993}
Cross, M.~C., Hohenberg, P.~C. Pattern formation outside of equilibrium. {\it Rev. Mod. Phys} {\bf 65}, 851-1112 (1993).
\bibitem{polymercollapsekinetics} Byrne, A., Kiernan, P., Green, D., Dawson, K.~A. (1995). Kinetics of homopolymer collapse. {\it J. Chem. Phys.} {\bf 1012}, 573-577 (1995).
\bibitem{Davide} Michieletto, D., Marenduzzo, D., Wani, A.~H. (2016) Chromosome-wide simulations uncover folding pathway and 3D organization of interphase chromosomes. arXiv:1604.03041.
\bibitem{NAR}  Brackley, C.~A., Johnson, J., Kelly, S., Cook, P.~R., Marenduzzo, D. (2016). Simulated binding of transcription factors to active and inactive regions folds human chromosomes into loops, rosettes and topological domains. {\it Nucl. Acids Res.} {\bf 44}, 3503-3512.
\end{thebibliography}

\newpage

\phantom{a}

\newpage

\appendix
\setcounter{figure}{0}
\makeatletter 
\renewcommand{\thefigure}{S\@arabic\c@figure}
\makeatother
	
\section*{Supplementary Information}
Here we give more details on the simulations (including parameter values), and on the continuum mean field model (derivation, linear stability analysis and amplitude equation); we also show additional results and figures which are discussed in the main text.
	
\section{Simulation Details}
The chromatin fiber is modeled as a bead-spring polymer with finitely-extensible non-linear elastic springs via a Kremer-Grest model~\cite{Kremer1990}. To map length scales from simulation to physical units, we can, e.g., set the diameter, $\sigma$, of each bead to $\sim 30$nm$\simeq 3$ kbp (assuming an underlying 30 nm fiber; of course, all our results would remain valid with a different mapping).

Letting  {$\bm{r}_i$} and  {$\bm{d_{i,j}} \equiv \bm{r}_j - \bm{r}_{i}$}  be respectively the position of the centre of the {$i$}-th bead and the vector of length  {$d_{i,j}$} between beads  {$i$} and  {$j$}, we can express the potential modeling the connectivity of the chain as
\begin{equation}
U_{\rm FENE}(i,i+1) = -\dfrac{k}{2} R_0^2 \ln \left[ 1 - \left( \dfrac{d_{i,i+1}}{R_0}\right)^2\right],  \notag
\end{equation}
for {$d_{i,i+1} < R_0$} and  {$U_{\rm FENE}(i,i+1) = \infty$}, otherwise; here we chose  {$R_0 = 1.6$ $\sigma$} and  {$k=30$}  {$\epsilon/\sigma^2$}. 

The bending rigidity of the chain is described through a standard Kratky-Porod potential, as follows
\begin{equation}
U_b(i,i+1,i+2) = \dfrac{k_BT l_p}{\sigma}\left[ 1 - \dfrac{\bm{d}_{i,i+1} \cdot \bm{d}_{i+1,i+2}}{d_{i,i+1}d_{i+1,i+2}} \right],\notag
\end{equation}
where we set the persistence length $l_p = 3 \sigma \simeq 90$ nm, which is reasonable for a chromatin fiber. 

The steric interaction between a chromatin bead, {$a$}, and a protein bridge, {$b$} (of sizes  {$\sigma_a=\sigma_b=\sigma$}), is modeled through a truncated and shifted Lennard-Jones potential  
\begin{equation}
U_{\rm LJ}(i,j) = 4 \epsilon_{ab} \left[ \left(\dfrac{\sigma}{d_{i,j}}\right)^{12} - \left(\dfrac{\sigma}{d_{i,j}}\right)^6 - \left(\dfrac{\sigma}{r_c}\right)^{12} + \left(\dfrac{\sigma}{r_c}\right)^6 \right], \notag 
\end{equation} 
for $d_{i,j}< r_c$ and 0 otherwise. This parameter, $r_c$, is the interaction cutoff; it is set to {$r_c=2^{1/6}\sigma$} for inactive proteins,  in order to model purely repulsive interactions, and to {$r_c=1.8\sigma$} for an active protein, so as to include attractive interactions. In both cases, the potential is shifted to zero at the cut-off in order to have a smooth curve and avoid singularities in the forces.  
Purely repulsive interactions, such as those between inactive proteins and chromatin segments, are modeled by setting $\epsilon_{ab}=k_BT$, while attractive interactions are modeled using: (i) $\epsilon_{ab}=3 k_BT$ (for non-specific interactions, Fig. 1); (ii) $\epsilon_{ab}=15 k_BT$ and $\epsilon_{ab}=4 k_BT$ (for non-specific and specific interactions respectively, Fig. 3); (iii) $\epsilon=4 k_BT$ (for non-specific interactions, Fig. 4); or (iv) as specified in Supporting Figure captions in other cases.

The total potential energy experienced by bead $i$ is given by
\begin{align}
	U_i&=\sum_j U_{\rm FENE}(i,j)\delta_{j,i+1} + \\ \notag
	&+\sum_j\sum_k U_{\rm b}(i,j,k)\delta_{j,i+1}\delta_{k,i+2} + \sum_j U_{\rm LJ}(i,j),
\end{align}
and its dynamics can be described by the Langevin equation
\begin{equation}
m \ddot{\bm{r}}_i = - \xi \dot{\bm{r}}_i - {\nabla U_i} + \bm{\eta}_i,
\label{langevin}
\end{equation}
where $m$ is the bead mass, $\xi$ is the friction coefficient, and $\bm{\eta}_i$ is a stochastic delta-correlated noise. The variance of each Cartesian component of the noise, $\sigma_{\eta}^2$, satisfies the usual fluctuation dissipation relation $\sigma_{\eta}^2 = 2 \xi k_B T$.

As is customary~\cite{Kremer1990}, we set $m/\xi = \tau_{\rm LJ}=\tau_{\rm B}$, with the LJ time $\tau_{\rm LJ} = \sigma \sqrt{m/\epsilon}$ and the Brownian time $\tau_{\rm B}=\sigma/D_b$, where $\epsilon$ is the simulation energy unit, equal to $k_BT$, and $D_b = k_BT/\xi$ is the diffusion coefficient of a bead of size $\sigma$. From the Stokes friction coefficient for spherical beads of diameter $\sigma$ we have that $\xi = 3 \pi \eta_{sol} \sigma$ where $\eta_{sol}$ is the solution viscosity. One can map this to physical units by setting the viscosity to that of the nucleoplasm, which ranges between $10-100$ cP, and by setting $T=300$ K and $\sigma=30$ nm, as above. From this it follows that $\tau_{\rm LJ} = \tau_{\rm B} = {3 \pi \eta_{sol} \sigma^3/\epsilon}\simeq 0.6-6$ ms; $\tau_B$ is our time simulation unit, used when measuring time in the figures in the main text and in this Supporting Information. 
The numerical integration of Eq.~\eqref{langevin} is performed using a standard velocity-Verlet algorithm with time step $\Delta t = 0.01 \tau_{\rm B}$  and is implemented in the LAMMPS engine. We perform simulations for up to $2\times10^5$ $\tau_{\rm B}$, which correspond to 2-20 minutes in real time. \\

\section{Phenomenological Mean Field Model for Bridges with Active Modification}
In our particle based simulations we observed the growth of clusters due to bridging interactions (see main text). 
When protein activation-inactivation reactions were absent, these clusters coarsened, resulting in one large macroscopic cluster in steady state. However, in the presence of these reactions, the clusters coarsened only up to a self-limiting size.
To better understand this transition from macrophase separation to microphase separation, and the involved length scales, we now develop a phenomenological minimal model for the dynamics of chromatin and proteins.
We describe the distribution of chromatin via the probability density field  $\rho({\bf x},t)$, and the density of active, or binding, and inactive, or non-binding, proteins by $\Phi_a({\bf x},t)\equiv \Phi({\bf x},t)$ and $\Phi_i({\bf x},t)$ respectively.

The starting point for our model is the free energy $\mathcal{F}=\int f({\bf x}){\rm d} {\bf x}$ where $f$ is the free energy density: 
\1
f=\4{D'_1}{2} \rho^2 + \4{D'_2}{2} \Phi^2 - \chi' \rho \Phi + \4{k'}{2} (\nabla \rho)^2 + \4{g'}{4}\rho^4. \2
Here, the first two terms describe diffusion of chromatin and proteins respectively, the third term describes the energy gain through bridging and the last two terms, multiplied by $k',g'$, respectively penalize sharp interfaces due to interfacial tension, and strong accumulations of chromatin due to short ranged repulsions.

Assuming diffusive dynamics here and using the fact that in the absence of protein modification, the number density of all species ($\rho, \Phi,\Phi_i$) is conserved, we can derive the equations of motions for our fields as done for model B dynamics~\cite{Chaikin2000}.
However, in the presence of active protein modification, we need an additional reaction term, so that our equations of motion read
\ea \dot \rho &=& -M_\rho \nabla^2 \4{\delta \mathcal{F}}{\delta \rho}, \label{eom1} \\
\dot \Phi_{a} &=& -M_{a} \nabla^2 \4{\delta \mathcal{F}}{\delta \Phi_{a}} - \alpha \Phi_a + \beta \Phi_i \label{eom2}.
\ee
Here $M_\rho$ and $M_{a}$ are dimensionless mobility coefficients of chromatin and activated proteins respectively, while $\alpha$ and $\beta$ are the activation and inactivation rates for proteins.
Since inactive proteins do not bind, we assume that they diffuse quickly, i.e. that their density field is uniform.

Now integrating Eq.~(\ref{eom2}) over the whole system and denoting the total number of active and inactive proteins with $N_a(t)$ and $N_i(t)$ respectively, 
we obtain $\dot N_a = -\alpha N_a + \beta N_i$. Conservation of the total protein number $N=N_a+N_b$ now yields 
$\dot N_i=(1+\beta/\alpha)N_i$ which approaches the steady state $N_i = \alpha N/(\alpha+\beta)$, i.e. $\Phi_i = \alpha/(\alpha+\beta)$, exponentially fast.
Now defining $\Phi_0:=(\beta/\alpha)\Phi_i=\beta/(\alpha+\beta)$ (and ignoring short-time effects due to possible `imbalances' between active and inactive proteins in the initial state), Eqs.~(\ref{eom1},\ref{eom2}) reduce to:
\ea
\dot \rho &=& M_\rho \nabla^2 [a_1 \rho - k \nabla^2 \rho - \chi \Phi + g \rho^3], \\
\dot \Phi &=& M_\Phi \nabla^2 [a_2 \Phi - \chi \rho] - \alpha (\Phi-\Phi_0),
\ee 
where for simplicity hereon we drop the subscript $a$ on $\Phi_a$ for active proteins. We also introduced $D_1=M_\rho a_1$ and $D_2=M_\Phi a_2$.

To further reduce these equations and to identify a minimal set of dimensionless control parameters, we now choose time and space units $t_u=1/\alpha$ and $x_u=\sqrt{D_2/\alpha}$ 
and redefine $\Phi=\Phi \chi M_\rho/D_2$. This leads to 
\ea
\dot \rho &=& \mathcal{D}_0 \nabla^2 \rho - A\nabla^4 \rho - \nabla^2 \phi + G \nabla^2 \rho^3, \label{rhod} \\
\dot \Phi &=& \nabla^2 \Phi - X \nabla^2 \rho - (\Phi-\Phi_0). \label{phid}
\ee
That is, our parameter space is spanned by the four dimensionless numbers $X=(\chi^2 M_\rho M_\Phi/D_2^2)$; $\mathcal{D}_0=(D_1/D_2)$; $A=\alpha k M_\rho/(D_2^2)$ and $G=gM_\rho/D_2$.

\section{Linear Stability Analysis}
To better understand in which parameter regimes we should expect (i) a uniform distribution of chromatin and proteins, (ii) cluster growth proceeding to macroscopic phase separation and (iii) microphase separation, we now perform a linear stability analysis. This analysis will equip us with a prediction for the self-limiting cluster size in regime (iii), matching the results of our particle based simulations.
We therefore study the response of the uniform phase to small perturbations in the density fields ($\rho,\Phi$). Linearising Eqs.~(\ref{rhod},\ref{phid}) around the uniform solution $(\rho,\Phi)=(\rho_0, \Phi_0)$, where $\rho_0$ is the DNA density as fixed by the initial state, leads to the following equations of motion for the fluctuations $\rho'=\rho-\rho_0, \Phi'=\Phi-\Phi_0$:
\ea
\dot \rho' &=& \mathcal{D}\nabla^2 \rho'- A\nabla^4 \rho - \nabla^2 \Phi', \label{lrhod} \\
\dot \Phi' &=& \nabla^2 \Phi' - X \nabla^2 \rho' - \Phi'. \label{lphid}
\ee
Here, we defined $\mathcal{D}:=\mathcal{D}_0+3G\rho^2_0$.
Fourier transforming Eqs.~(\ref{lrhod},\ref{lphid}) and using $Q:={\bf q}^2$ leads to the following dispersion relation (or characteristic polynomial),
\begin{align} 
\lambda(Q)&=\4{1}{2}\left[-1 -Q\left(1+\mathcal{D}+A Q\right) \right. \pm \notag \\ 
& \left. \pm \sqrt{\left[-1 + Q(\mathcal{D}-1+AQ)\right]^2 + 4 Q^2 X} \right], \label{dispr} 
\end{align}
which links the growth rate $\lambda$ of the fluctuation with its wavevector $Q$. An analysis of this relation leads us to the instability criterion
\1 \sqrt{X}>\sqrt{X_C}:=\sqrt{A}+\sqrt{\mathcal{D}},\2
which translates, in physical units, to 
\1 \chi>\sqrt{\frac{k \alpha}{M_\Phi}}+\sqrt{\frac{D_2}{M_\Phi}\left[\frac{D_1}{M_\rho} +3g \rho_0^2\right]}.\2
This criterion determines the transition line (hypersurface) between regions (i) and (ii/iii) in the parameter space. Hence, if the bridging interactions are sufficiently large, small fluctuations around the uniform state will grow to form clusters. 
Remarkably, this instability and the corresponding emergence of order (clustered phase) is not contingent on the presence of a certain minimal protein (or DNA) density, suggesting that even a very low protein concentration is sufficient to trigger clustering.

To map out the transition line from macrophase separation to microseparation (at the onset of instability), it is useful to consider the wavelength at which instability first occurs. 
From Eq.~(\ref{dispr}) and $q_c = \left. \partial_q \lambda(q)=0\right|_{X=(\sqrt{A}+\sqrt{D})^2}$, we find $q_c = (D/A)^{1/4}$, corresponding, in physical units, to the length scale 
\1 L_c = \4{2\pi}{q_c} = 2\pi \left(\4{D_1 D_2 + 3M_\rho g\rho_0^2 D_2}{\alpha k M_\rho} \right)^{1/4}. \2
Thus, in an infinite system, coarsening only occurs for $\alpha=0$. [In finite systems macrophase separation is observed if $\alpha$ is small enough that $L_c$ exceeds the system size.]
From this analysis we expect the average particle number per cluster to scale as $N \propto L_c^3 \propto \alpha^{-3/4}$ (at least close to the onset of instability). This value agrees well with the numerically observed scaling of $N \propto \alpha^{-0.76}$ (Fig.~S1), supporting the view that the essential physics of chromatin clustering can be described and understood within our simplified mean field theory. 

For completeness, we also calculate the boundaries of the instability band from Eq.~(\ref{dispr}), which, after translating back into physical units (for $M_1=M_2=1$), read as follows: 
\ea 
q_\pm &=&\4{1}{\sqrt{2 D_2 K}} \sqrt{\nu \pm \sqrt{\nu^2 - 4 D_2 K \alpha (D_1 + 3 g \rho_0^2)}}, \\
\nu &=& \left[\chi^2 - D_1 D_2 - 3D_2 g \rho_0^2 - K \alpha \right].
\ee
At the onset of instability, we find $\nu \rightarrow 4 D_2 K \alpha (D_1 + 3 g \rho_0)$ and hence we recover the $\alpha^{1/4}$-scaling of the onset mode.
In contrast, the boundaries of the instability band scale in a more complicated way which is nonuniversal in $\alpha$. 


\begin{figure}
\begin{center}
\includegraphics[width=0.48\textwidth]{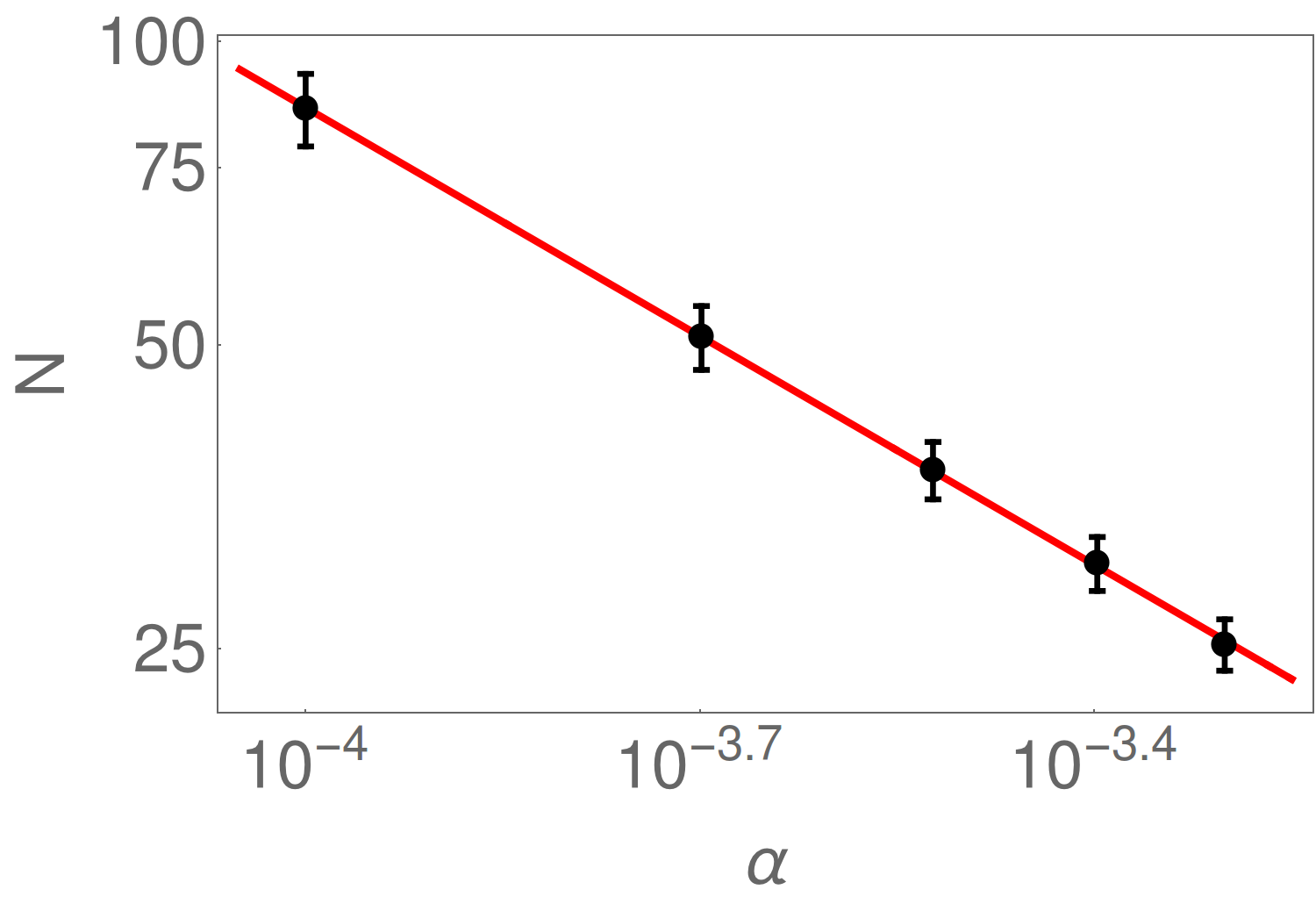}
\caption{\textbf {Comparision between mean field predictions and simulation results for cluster size.} Simulations are like the ones illustrated in Fig. 1 of the main text (except for the value of $\alpha$, which is varied). Dots show saturation values ($\pm$ SD) of number of particles per cluster $N$ (after $1.5\times 10^5$ simulation units); the line shows a least-squares fit with a slope of $-0.756$.}
\end{center}
\label{clscaling}
\end{figure}

\section{Amplitude Equations}
We now perform a perturbative analysis of the linearly unstable modes (fluctuations) close to onset of instability. 
This analysis will lead us to a further reduced effective model, describing the linear growth and nonlinear saturation of chromatin clusters on large scales and at long timescales. 

We begin by rewriting Eqs.~(\ref{eom1},\ref{eom2}) as 
\1 \mathcal{L} \left(\begin{matrix} \rho' \\ \Phi' \end{matrix}\right) + \mathcal{N}- \left(\begin{matrix} \dot \rho' \\ \dot \Phi' \end{matrix}\right) =  \left(\begin{matrix} 0 \\ 0 \end{matrix}\right), \2
where the linear operator $\mathcal{L}$ and the nonlinear term $\mathcal{N}$ represent
\1 \mathcal{L}=\left(\begin{matrix} 
\mathcal{D} \partial^2_x - A \partial^4_x & - \partial^2_x \\                
 -X \partial_x^2 & \partial_x^2 -1              \end{matrix}\right)
;\quad 
\mathcal{N}=G\left(\begin{matrix} \partial_x^2 \rho'^3 + 3 \rho_0 \partial_x^2 \rho'^2 \\ 0 \end{matrix}\right).
\label{reom} \2
Now, we replace (as usual, see \cite{Cross1993}) 
\1 X \rightarrow (1+\epsilon) X_C;\quad \partial_x \rightarrow \partial_x + \sqrt{\epsilon}\partial_X; \quad \partial_t \rightarrow \epsilon \partial_T \2
where $\epsilon=(X-X_C)/X_C$ and expand the fields as
\1 
\rho' = \Sum{n=1}{\infty} \epsilon^{n/2} \rho_{n-1};\quad \Phi'=\Sum{n=1}{\infty} \epsilon^{n/2} \Phi_{n-1}.
\2
 
Next, we plug these expansions into Eqs.~(\ref{reom}) and solve the resulting equations to lowest order ($\epsilon^{1/2}$). Using the Ansatz $\rho_0 = \mathcal{A}\exp{(\I q_c x)} + c.c.$ and $\Phi_0=\mathcal{A}_\phi \exp{(\I q_c x)} + c.c.$ with amplitudes $\mathcal{A},\mathcal{A}_\phi$, we find $q_c=(\mathcal{D}/A)^{1/4}$ reproducing the corresponding result from our linear stability analysis (see above), as well as $\mathcal{A}_\phi=\mathcal{A}(\mathcal{D}+ A q_c^2)=\mathcal{A} \sqrt{\mathcal{D}X_C}$ which fixes the relation between the amplitudes of both density fields. 
The solution of our perturbative equations to order $\epsilon^{1/2}$ then reads $\rho'=2\mathcal{A}\cos{q_c x}$ with the so-far unknown amplitude $\mathcal{A}$.

The result to order $\epsilon$ turns out not to be particular useful for our purpose, as solving it would provide us with a similar result as to order $\epsilon^{1/2}$), but with another unknown amplitude $\mathcal{A'}$ yielding a higher order correction to the solution $\rho'=2\mathcal{A}\cos{q_c x}$.
Since we are looking only for the lowest order result in $\epsilon$ we directly consider the perturbative equations of motion to order $\epsilon^{3/2}$.
As usual \cite{Cross1993}, we do not attempt to solve the corresponding equations explicitly, but apply Fredholm's theorem providing solvability conditions, which determine an equation of motion for $\mathcal{A}$. 
After a long but straightforward calculation and transforming back to coordinates $t,x$ we find:
\1 c_t \dot{\mathcal{A}} = \epsilon \mathcal{A} + c_x \partial_x^2 \mathcal{A} + c_3 \mathcal{A}^3 \label{rgle}, \2
where 
\ea c_t &=& \sqrt{\4{A}{X_C}}\left(1+\4{1}{\mathcal{D}}\right), \label{ct} \\
c_x &=& \4{4A}{\sqrt{\mathcal{D} X_C}}, \label{cx} \\
c_3 &=& \4{3G}{\sqrt{\mathcal{D} X_C}}. \label{c3}
\ee
Eq. (\ref{rgle}) is a variant of the real Ginzburg-Landau equation, here describing, together with the coefficients Eqs.~(\ref{ct}--\ref{c3}), the dynamics of chromatin and proteins close to the onset of instability. 
In this equation $\epsilon/(c_t t_u)$ is the initial growth rate of protein clusters; $x_u \sqrt{\epsilon/c_3}$ describes the amplitude of their saturation (related to their density) 
for a given $X>\left[\sqrt{A}+\sqrt{\mathcal{D}}\right]^2$
and $x_u \sqrt{c_x}$ is a correlation length, describing a scale of spatial modulations of the saturation amplitude of DNA clusters. 

Although we equipped our original equilibrium model with reaction terms which drive it out of equilibrium, its large scale and long time dynamics (i.e., Eq.~(\ref{rgle})) can be effectively mapped (at least close to onset of instability) onto a potential system with the following Lyapunov functional: 
\1 \mathcal{V}[\mathcal{A}]= \int {\rm d} x \left[ -\epsilon |\mathcal{A}|^2 + \4{c_3}{2}|\mathcal{A}|^4 + c_x^2 |\partial_X \mathcal{A}|^2 \right],\2 
\1 \dot{\mathcal{A}}= \4{-1}{c_T} \4{\delta V}{\delta \mathcal{A}}. \2
Hence, quite remarkably, the dynamics of the present reaction-diffusion system can be mapped, within this approximation, onto a system which is purely relaxational.

\begin{figure}[!th]
\begin{center}
\includegraphics[width=1.1\columnwidth]{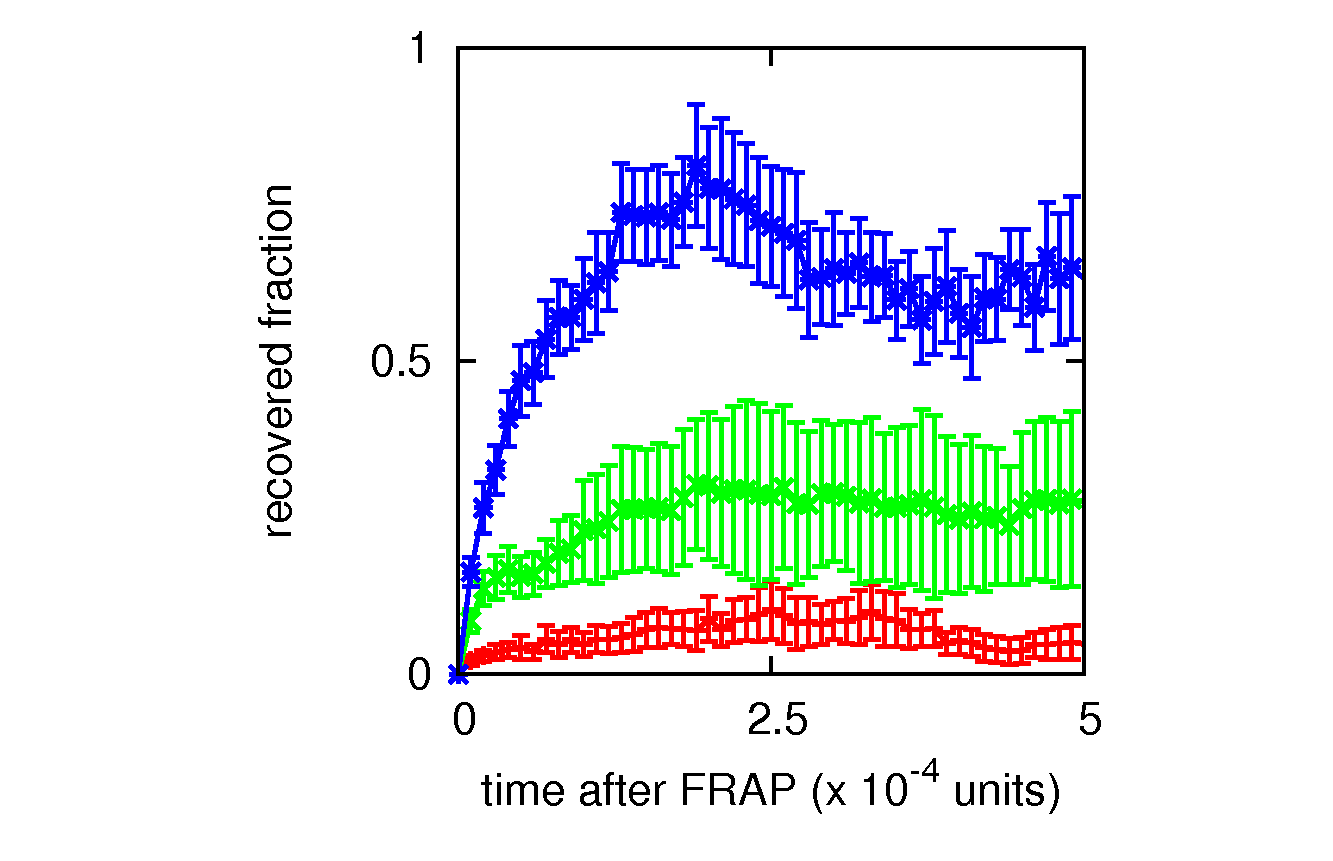}
\caption{\textbf{Comparison of FRAP recovery for non-switching and switching proteins}. FRAP recovery, measured as the number of unbleached proteins which are in the bleached volume after bleaching. The signals are normalized with the number of proteins in the bleached volume at the time of bleaching. As in Fig. 3Bi, the bleached volume is a sphere of size $50 \sigma$. Error bars give SD of mean, and time is given in multiples of $10^4$ simulation units. Values of the specific and non-specific interactions, and of $\alpha$, were respectively: $15 k_BT$, $4 k_BT$, $0$ (red curve), $8 k_BT$, $3 k_BT$, $0$ (green curve), and $15 k_BT$, $4 k_BT$, $0.0001$ inverse Brownian times (blue curve). It can be seen that varying the values of non-specific and specific interactions can lead to FRAP recovery also for $\alpha=0$ (green curve), although, in the absence of fine tuning, this is to a smaller extent with respect to $\alpha\ne 0$ (blue curve).}
\end{center}
\end{figure}

\section{Additional simulation results}

In this section we present additional simulation results, which complement those discussed in the main text.

Figure S2 shows that the FRAP signal (following simulated photobleaching of a spherical spot of size 50 $\sigma$) shows recovery also for equilibrium bridges, if the specific and non-specific interactions are carefully tuned. However, protein modification provides a more robust way to achieve this, which simultaneously allows stable binding (when the protein is in the active state), and fast turnover (due to the unbinding and diffusion of inactive proteins).

Figure S3 shows the cluster size for different parameter values for the case of non-specific protein-chromatin interactions. This demonstrates that it can be varied significantly (by about an order of magnitude), and is particularly sensitive to the protein concentration.

\begin{figure}[!h]
\renewcommand{\thesubfigure}{A}
\subfloat[]{\includegraphics[width = 0.9\columnwidth]{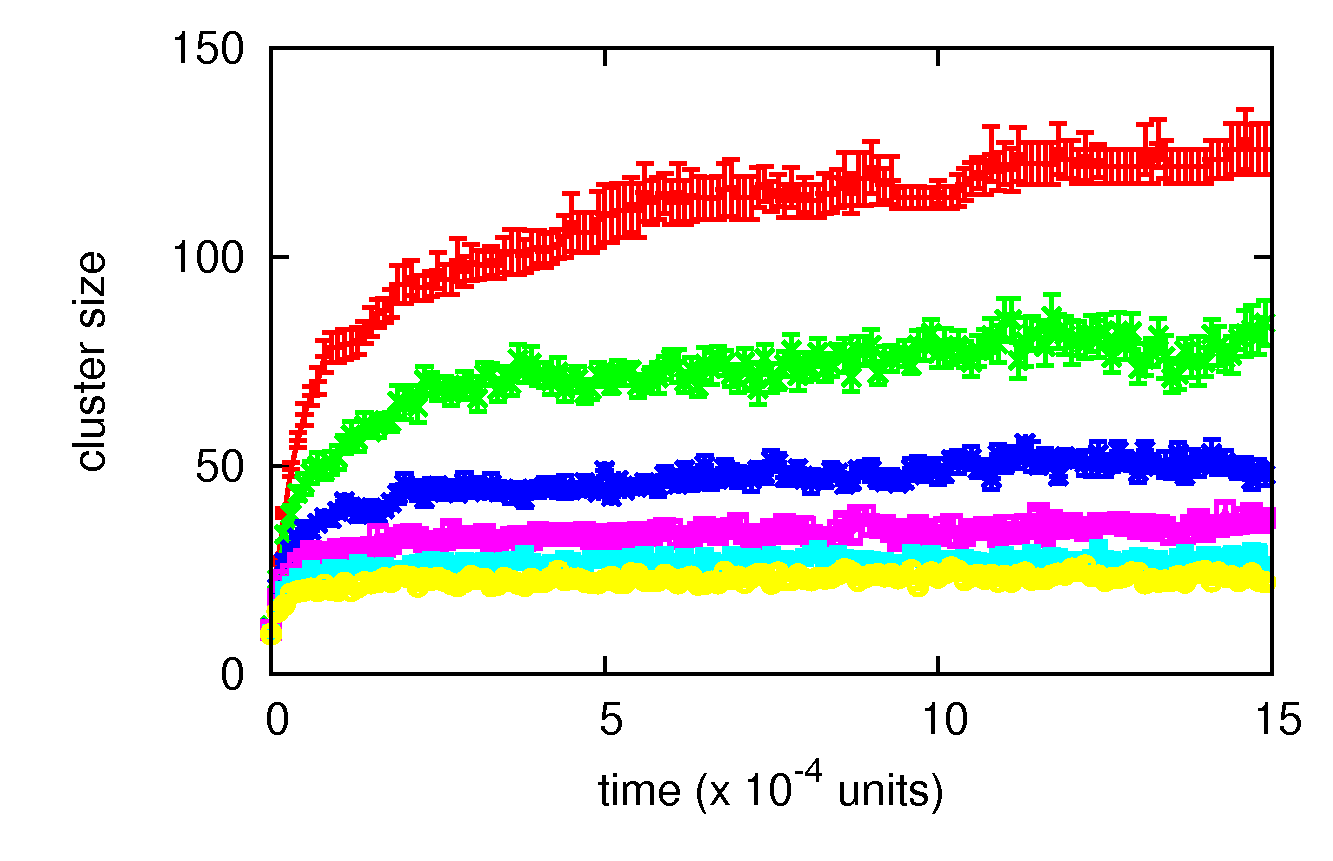}}
\renewcommand{\thesubfigure}{B}
\subfloat[]{\includegraphics[width = 0.9\columnwidth]{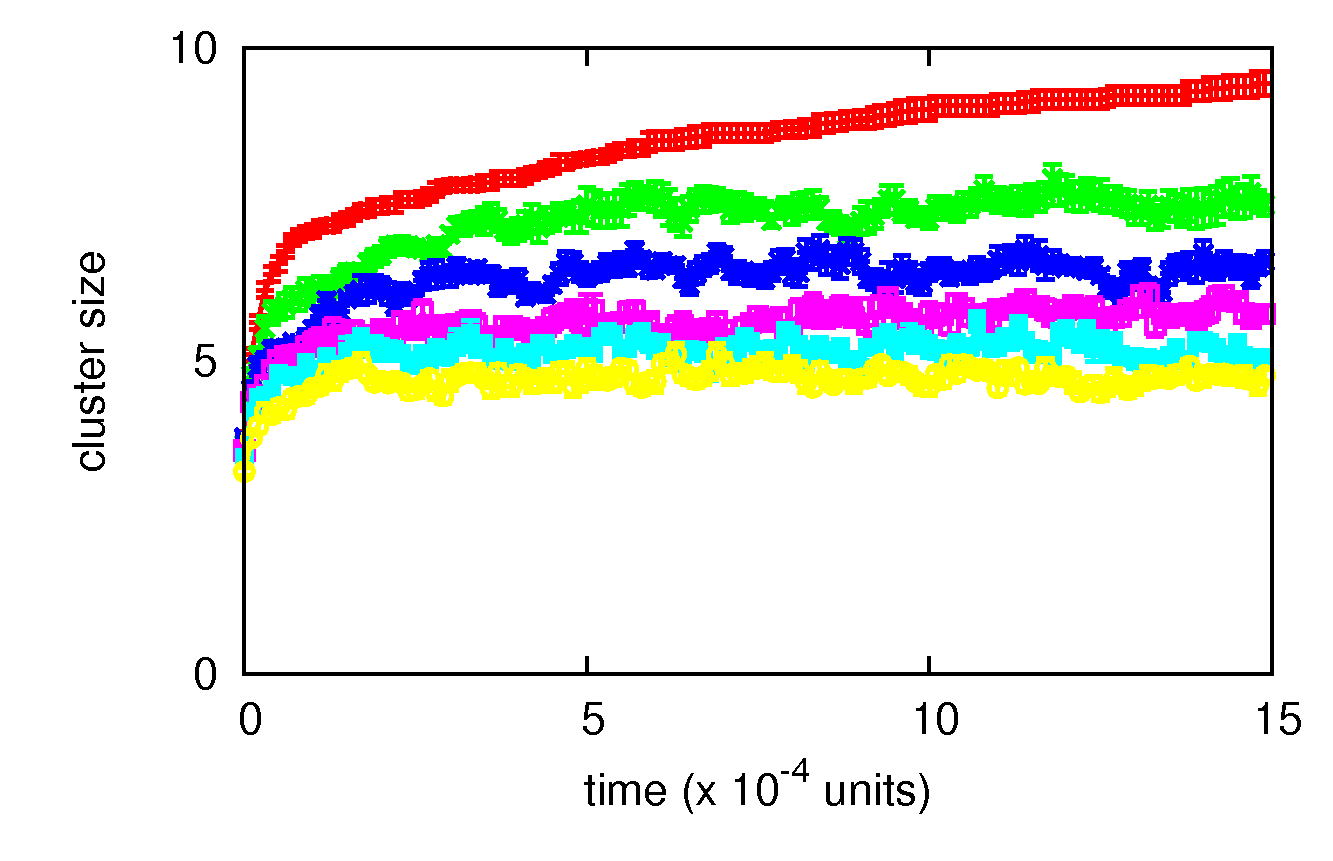}}
\caption{\textbf{Cluster size with specific binding.} (A) Plot of the average number of proteins in a cluster versus time ($\pm$ SD), for $N=2000$ switching proteins binding to the chromatin fiber, both specifically (interaction strength $15$ $k_BT$, cut-off $1.8 \sigma$), to every 20-th bead in the polymer, and non-specifically (interaction strength $4$ $k_BT$, cut-off $1.8 \sigma$) to any other bead. From top to bottom, curves correspond to $\alpha=0$ (in which case half of the proteins are non-binding, and half binding), $0.0001$, $0.0002$, $0.0003$, $0.0004$, $0.0005$ respectively. (B) Same as (A), but now for $N=500$ switching proteins, with specific interaction strength of $8 k_BT$ and non-specific interaction of $3 k_BT$; the interaction cut-off is $1.8 \sigma$.}
\end{figure}

\begin{figure*}[!th]
\renewcommand{\thesubfigure}{Ai}
\subfloat[]{\includegraphics[width = .85\columnwidth]{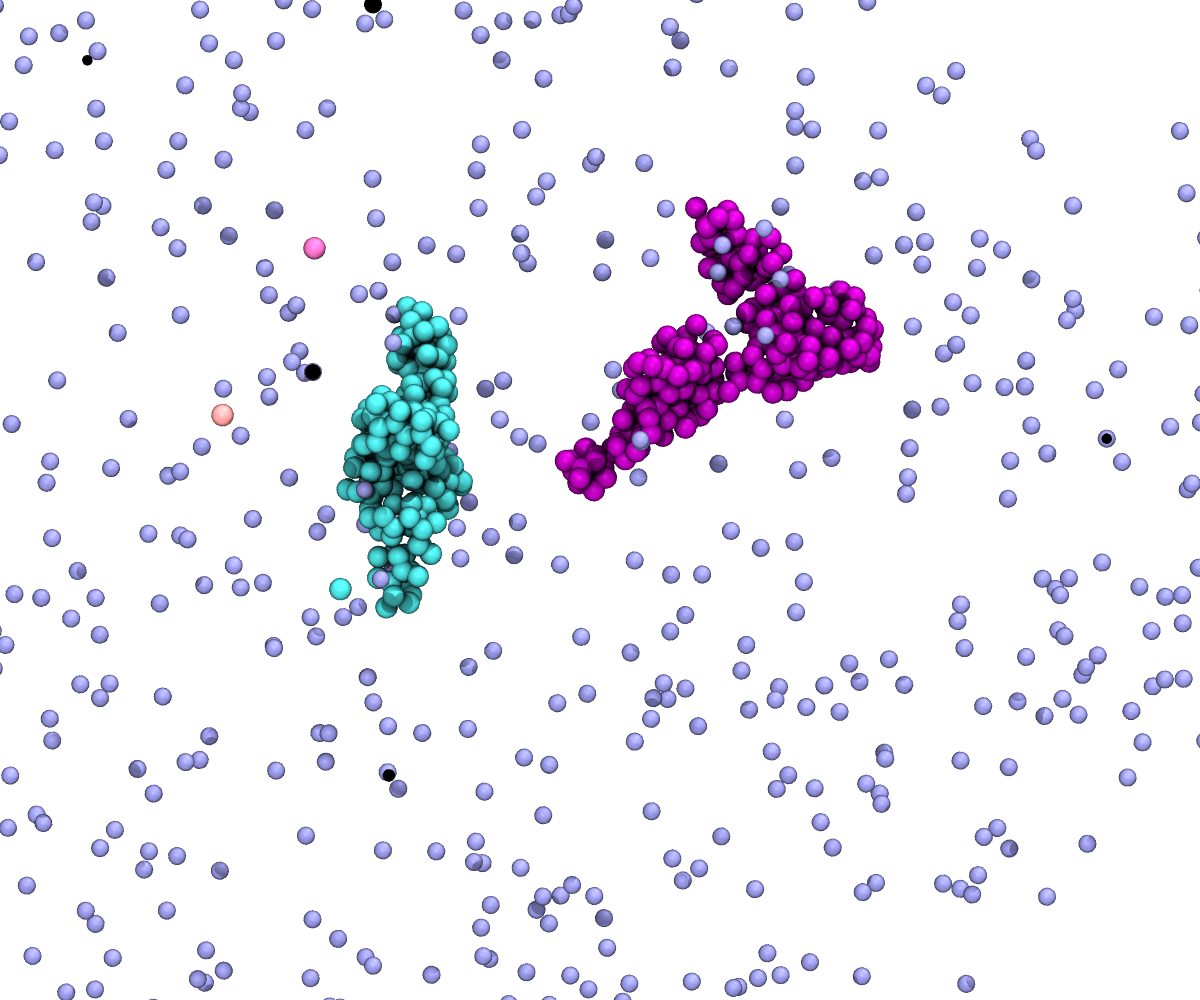}}\quad
\renewcommand{\thesubfigure}{Bi}
\subfloat[]{\includegraphics[width = .85\columnwidth]{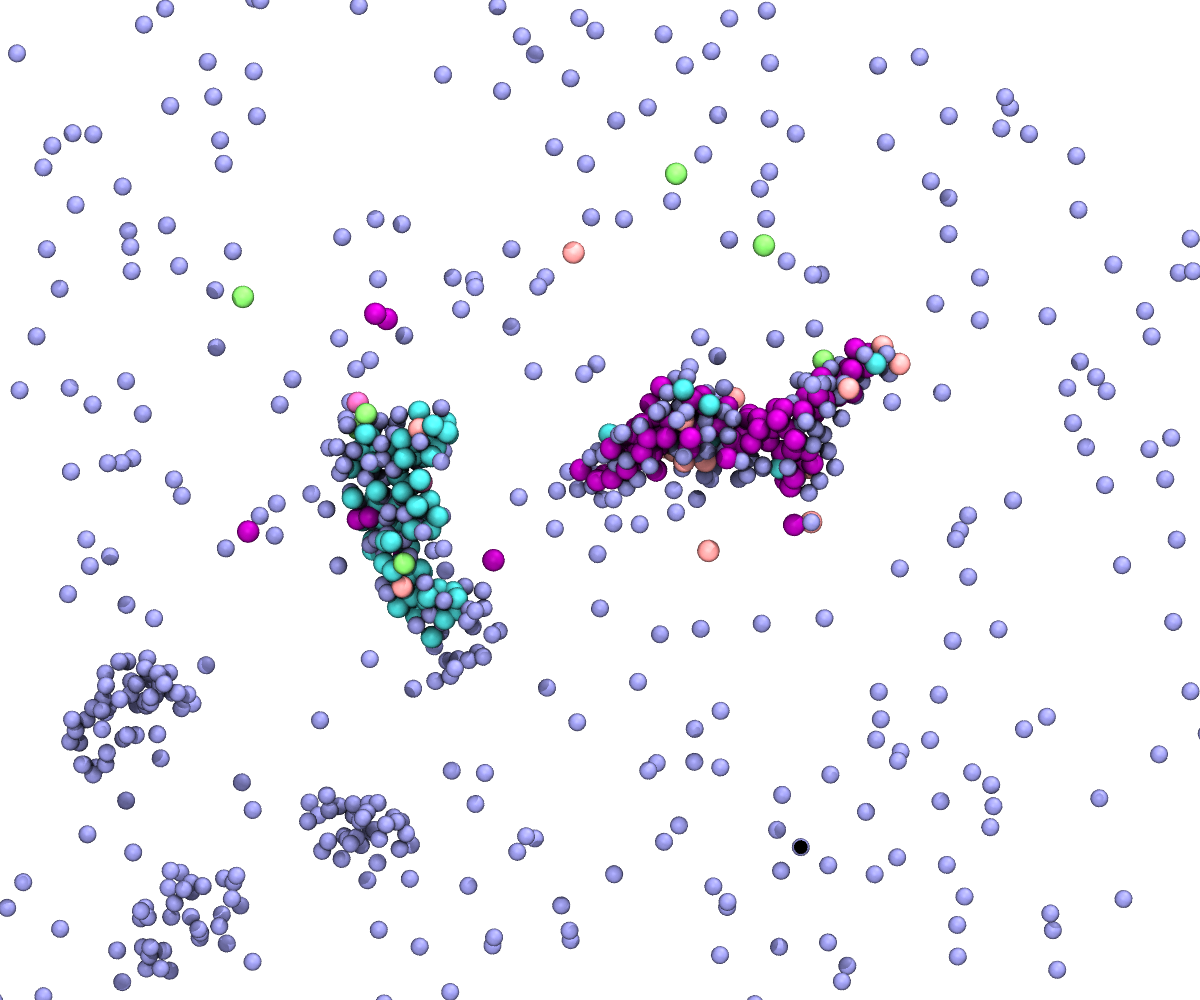}}\\
\renewcommand{\thesubfigure}{Aii}
\subfloat[]{\includegraphics[width = .85\columnwidth]{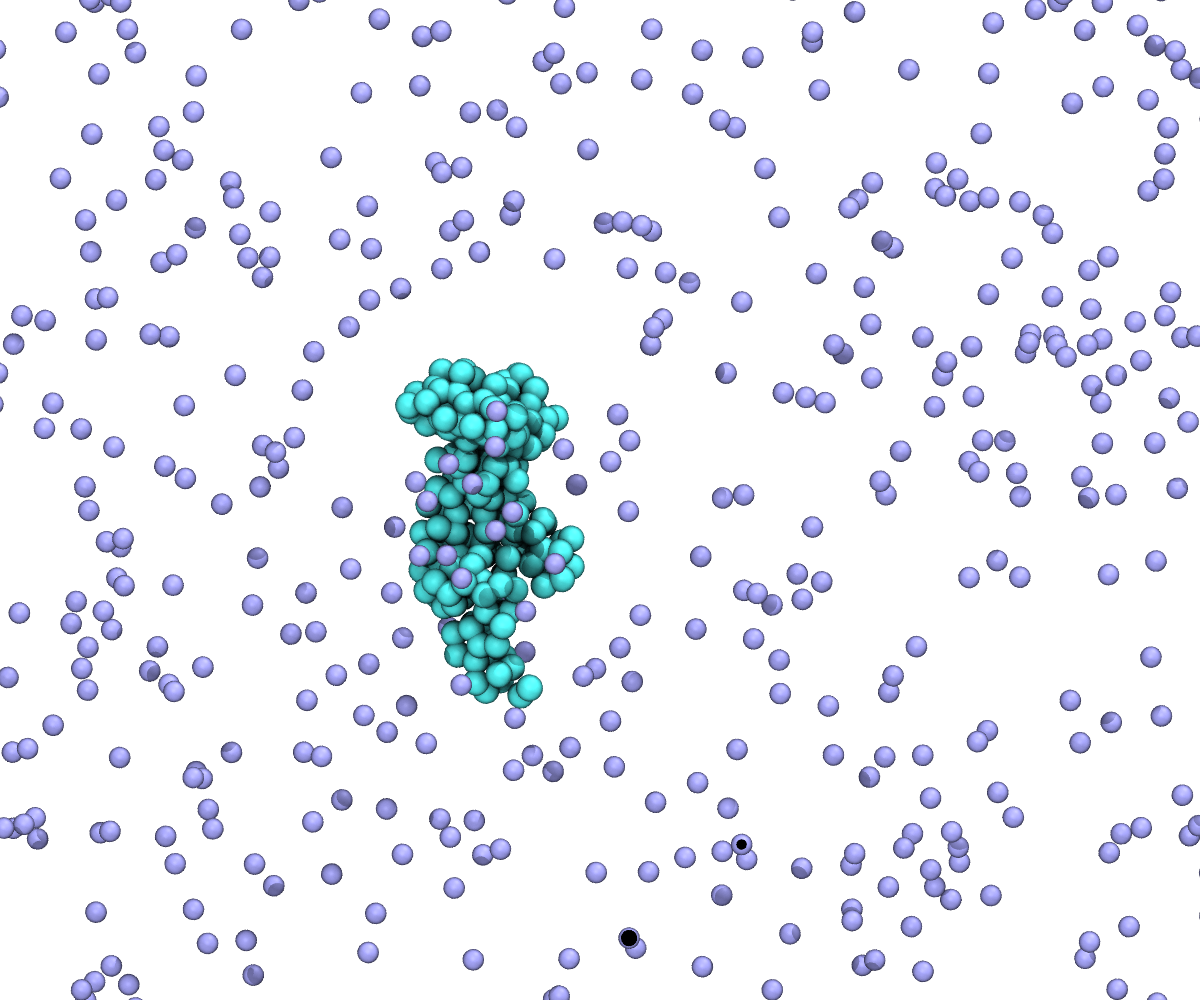}}\quad
\renewcommand{\thesubfigure}{Bii}
\subfloat[]{\includegraphics[width = .85\columnwidth]{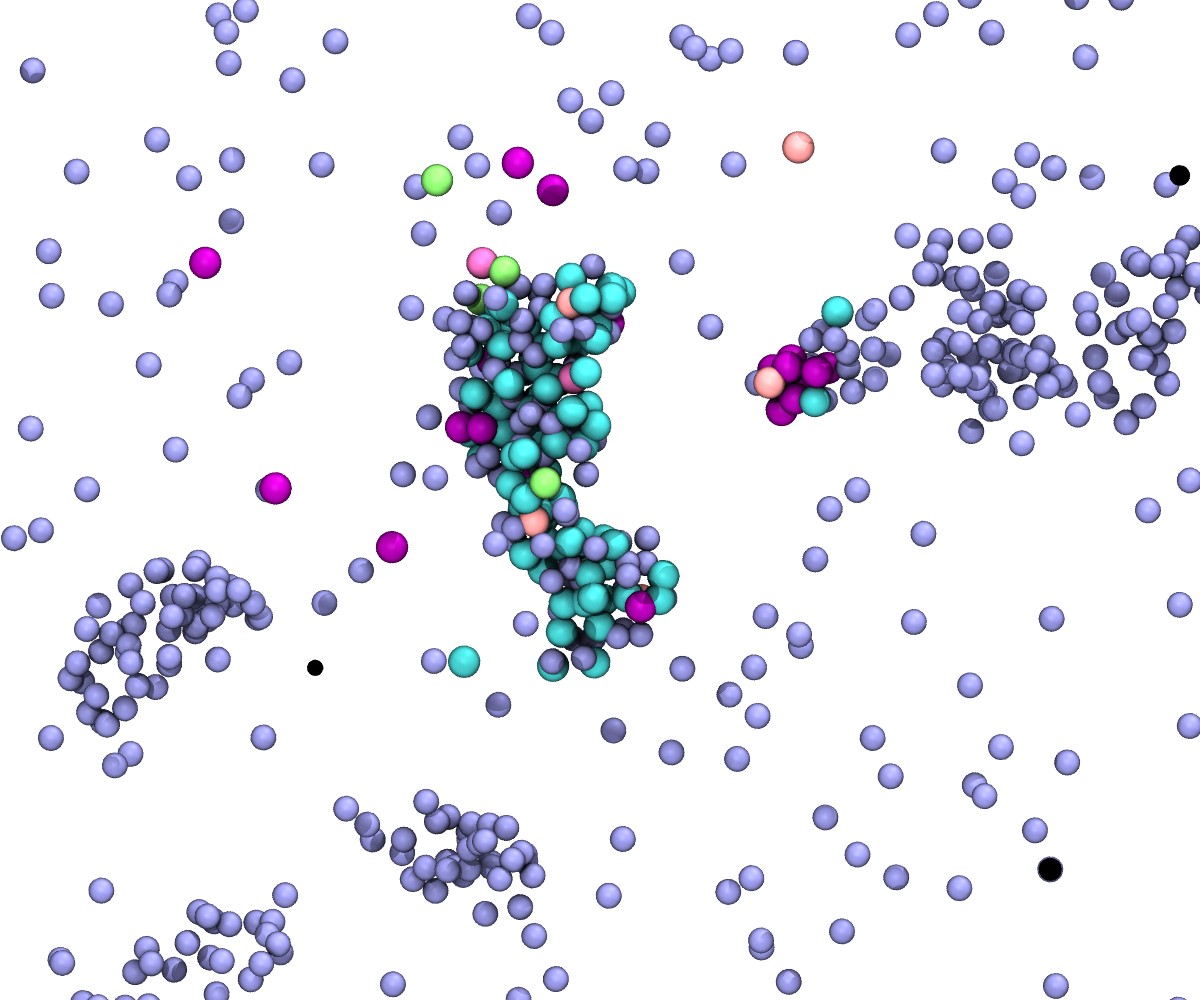}}\\
\caption{\textbf{Switching proteins form clusters which retain memory of their shape.} This figure follows the evolution of clusters in a simulation analogous to that of Fig. 3A in the main text; the same parameters apply. Only proteins -- and not chromatin beads -- are shown for clarity. (A) Snapshots taken $10^4$ time units after equilibration, for non-switching proteins, showing two clusters (beads are colored according to the cluster they belong to); (ii) shows another cluster. (B) Snapshots of the same regions shown in (A) after another $10^5$ simulation units, and after allowing the proteins to now switch ($\alpha=0.0001 \tau_{\rm B}^{-1}$). Clusters recycle their constituent proteins whilst retaining a very similar shape.}
\end{figure*}

\begin{figure}[!th]
\begin{center}
\includegraphics[width = 1\columnwidth]{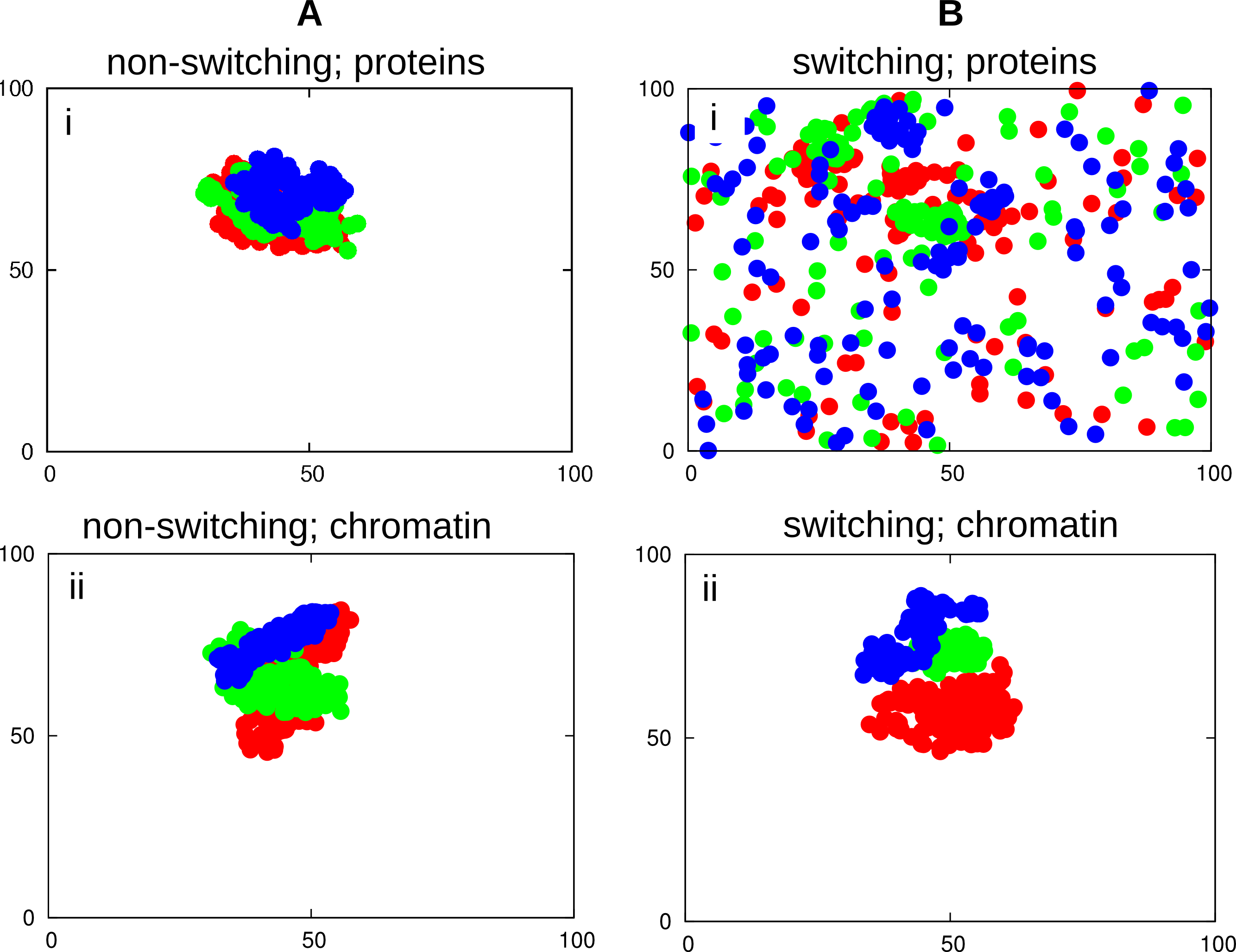}
\end{center}
\caption{\textbf{Trajectories of proteins and high-affinity chromatin beads.} Simulations are as in Fig. 3 of the main text; the same parameters apply. Positions of proteins and chromatin beads are shown in a 2D projection of the simulation domain, positions on the axes are measured in units of $\sigma$. (A) Non-switching proteins. (i) Red, green and blue circles denote positions of three non-switching proteins, recorded every $100$ $\tau_{\rm B}$ in a simulation (total length $1.5 \times 10^5$ simulation units. In this case, all three proteins remain bound to one cluster throughout the time series. (ii) Red, green and blue circles denote positions of three high affinity chromatin beads, again recorded every $100$ $\tau_{\rm B}$ in the same simulation. All three chromatin beads remain in the same cluster. (B) Same as (A), but for switching proteins ($\alpha=0.0001$ inverse Brownian times). Now the three switching proteins diffuse through the whole space, while the three chromatin beads are still confined; this shows that the underlying scaffold of the cluster persists as the proteins are recycled.}
\end{figure}

\begin{figure}[!h]
\begin{center}
\includegraphics[width = 1\columnwidth]{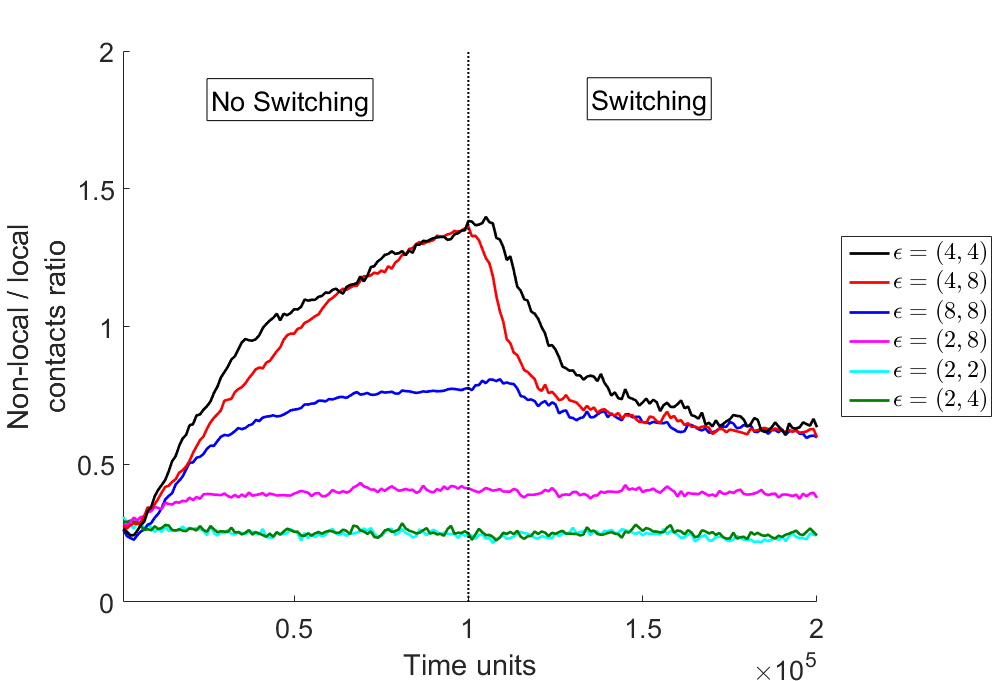}
\end{center}
\caption{\textbf{Chromatin contacts for different choices of specific and non-specific binding affinities.} The plot shows the fraction of non-local versus local contacts for a chromatin fiber; fiber patterning and all parameters are as in Fig. 4 of the main text. Simulations initially involved non-switching proteins; half-way through the simulation, proteins began to switch ($\alpha=0.0001$ inverse Brownian times). Contacts are classified as local (non-local) if they involve beads separated less than (more than) $400$ beads along the chain (or $1.2$ $Mbp$. Non-specific ($\epsilon_1$) and specific ($\epsilon_2$ interaction energies are indicated on the right of the plot, in the format $(\epsilon_1,\epsilon_2)$.}
\end{figure}

Figs. S4 and S5 highlight some further properties of the recycling clusters. In particular, Fig. S4 shows that these clusters retain memory of their shape even as the proteins which constitute them change. Figure S5 shows the dynamics of some protein and chromatin beads with and without modification. Without modification, once proteins bind to a cluster they diffuse little for the rest of the simulation, whereas with modification they sample the whole simulation domain. Contrary to this, the dynamics of the chromatin beads within a cluster is similar with and without modification: they diffuse very little. This explains why clusters keep their shape: while proteins bind and unbind, the underlying chromatin backbone is largely unchanged.

Finally, Fig. S6 shows how the effect of protein switching on the ratio between non-local and local contacts, shown in Fig. 4 in the main text, is affected by the values of non-specific and specific interactions. \\

\newpage

\phantom{a}
\newpage

\section{Timescale estimates for FRAP experiments and TAD folding}

Here we provide a series of simple estimates for the value of the relevant timescales in our problems. Consider first a fluorescence-recovery-after-photobleaching, or FRAP, experiment, where a cluster of size $\sigma_{\rm cl}\sim 0.1-1$ $\mu$m is inside the bleached spot, which we imagine has a diameter of $\sigma_{\rm FRAP}\sim 1$ $\mu$m. In this Section, as previously, $\sigma$ will instead denote the size of a typical chromatin-binding complex, or chromatin bead (as previously, we imagine this is $\sim 30$ nm).

What is the timescale for the recovery of the FRAP signal? Clearly, this depends on the underlying dynamics of the bleached/unbleached proteins.  If proteins diffuse freely, then unbleached proteins can enter the bleach spot to give recovery within a time, $\tau_{\rm diff}$, proportional to
\begin{equation}
\tau_{\rm diff} \sim \frac{\sigma_{\rm FRAP}^2}{D}.
\end{equation}
For a protein size $\sigma\sim 30$ nm, and if the nucleoplasm viscosity is $10$ cP (ten times that of water), the diffusion coefficient is $\sim 1.4 \mu$m${^2}$ s${-1}$, so that $\tau_{\rm diff}\sim 1$s, which is too fast to account for FRAP response of nuclear bodies (furthermore, of course, freely diffusing proteins could not self-organise into clusters).

If, instead, non-switching binding proteins create a cluster, then the FRAP signal recovers when some proteins unbind, and others replace these from the soluble (unbleached) pool. As the former process is slower than the latter (which relies again on diffusion), we can equate the FRAP recovery timescale to the time needed for an equilibrium protein to unbind from the cluster, which can be estimated as,
\begin{equation}
\tau_{\rm non-switch} \sim \frac{\sigma_{\rm cl}^2}{D}\exp{\left(\frac{\Delta U}{k_BT}\right)}
\end{equation}
where $\Delta U$ indicates the strength of chromatin-protein interaction. If we assume an interaction of $10$ kcal/mol, consistent with either multiple non-specific or a single specific DNA-protein interactions, then $\tau_{\rm non-switch} > 10^5$ s, which is too slow to account for the FRAP recovery observed in nuclear bodies. Clearly, changing $\Delta U$ will change $\tau_{\rm non-switch}$, but in order for the estimate to be in the observed range, the interaction energy would have to be finely tuned, and would be significantly lower than that seen in typical DNA-protein interactions. 

If, finally, switching proteins are in the cluster, then the unbinding time, which again can be equated to the FRAP recovery time, is simply
\begin{equation}
\tau_{\rm non-switch} \sim \alpha^{-1}.
\end{equation}
For typical post-translational modification, or transcription termination, this is in the several seconds to minutes timescale, which is compatible with experimental results. 

Aside from FRAP, another important timescale is that over which local TADs form (e.g., in Fig. 4), $\tau_{\rm TAD}$. In analogy with polymer collapse and heteropolymer folding (see, e.g., Ref.~\cite{polymercollapsekinetics}), we expect $\tau_{\rm TAD}$, to be a power law in the number of monomers in the TAD, say $M$, where the prefactor should describe microscopic (diffusion) dynamics of a monomer. Dimensional analysis then suggests 
\begin{equation}
\tau_{\rm TAD} \sim \frac{\sigma^2}{D} M^z \sim \tau_{\rm B} M^z
\end{equation}
where $z$ is a scaling exponent. The Brownian time $\tau_{\rm B}$ is of the order of $10^{-3}$ s with previous assumptions for viscosity and monomer size, while in our simulations $z\simeq 1$ at least up to $M\sim 100$ (corresponding to 300 kbp). Also for eukaryotic chromosomes, TAD size is between 100 kbp and 1 Mbp, so $M$ is at most a few hundred. Therefore, if $z=1$, we estimate $\tau_{\rm TAD}$ to be of the order of $1$ s, smaller than typical modification times -- even assuming a larger effective value of $z$ (e.g., $z=2$ gives at most $\tau_{\rm TAD}$ of order of $1$ min). Previous large-scale simulations also confirm that eukaryotic TADs form in minutes~\cite{Davide,NAR}. These estimates explain why switching proteins in our simulations can still form TADs in pretty much the same way as non-switching proteins, and suggest that the same should also hold for real chromosomes.

\section{Captions of Supplementary Movies}

{\bf Supplementary Movie 1:} A movie of the simulation shown in Figure 1C of the main text. Proteins do not switch ($\alpha=0$). First a snapshot $10^4$ simulation units after equilibration is shown: a number of small clusters have formed. Then the subsequent dynamics are shown: clusters grow and merge, and coarsening proceeds indefinitely. \\

{\bf Supplementary Movie 2:} A movie of the simulution shown in Figure 1D of the main text. Proteins switch at a rate $\alpha=0.0001$ inverse Brownian times. Switching arrests coarsening, and leads to clusters of self-limiting size in steady state.\\

{\bf Supplementary Movie 3:} Parameters for this Movie are as in Fig. 3 of the main text for the $\alpha=0$ case. Chromatin beads are not shown for simplicity. The movie starts with clusters which have formed during $10^4$ simulation units following equilibration. The proteins are colored according to the cluster they belong to when the movie starts; proteins not in any clusters at that time are gray. The movie then follows the dynamics with non-switching proteins, for another $10^4$ simulation units: it can be seen that colored clusters persist, therefore photobleaching such a cluster would lead to little or no recovery of signal in the cluster. \\

{\bf Supplementary Movie 4:} As Supplementary Movie 3, but now with switching proteins ($\alpha=0.0001$ inverse Brownian times). Proteins are colored according to the initial clusters; by the end of the simulations all clusters have mixed colors. While proteins in clusters recycle, the cluster retains the same overall shape. \\

{\bf Supplementary Movie 5:} As in Supplementary Movie 4, but a zoom on two clusters to show more clearly clusters retain a ``memory'' of their shape.\\

{\bf Supplementary Movie 6:} A movie of the simulation shown in Figure 4 in the main text. The first half of the simulation involves non-switching proteins and lasts $10^5$ simulation units: two clusters form. Proteins are black; yellow chromatin beads are binding, while blue ones are non-binding. During the second half, proteins are able to switch ($\alpha= 0.0001$ inverse Brownian times); clusters split and interdomain interactions are suppressed.

\end{document}